\documentclass{aa}
\usepackage{amssymb}
 \usepackage{booktabs}
\usepackage{color}
\usepackage[T1]{fontenc}
\usepackage[utf8]{inputenc}

\usepackage{savesym}
\usepackage{multirow}
\savesymbol{tablenum}
\usepackage{siunitx}
\usepackage{booktabs}
\usepackage{ragged2e}
\usepackage[inkscapeformat=png]{svg}
\usepackage{subfigure}
\usepackage{graphicx}
\usepackage{float}
\usepackage{txfonts}
\usepackage{hyperref}
\hypersetup{
    colorlinks=true,
    linkcolor=blue,
    citecolor=blue,  
    urlcolor=blue,
    }

\begin{document}

% \title{Cherenkov image correction method for cloud-affected observations}
\title{A novel image correction method for cloud-affected observations with Imaging Atmospheric Cherenkov Telescopes}
\titlerunning{A novel image correction method for cloud-affected observations with IACTs}
   \author{Natalia Żywucka\inst{1}
          \and
          Julian Sitarek\inst{1}
          \and
          Dorota Sobczyńska\inst{1}
          \and
          Mario Pecimotika\inst{2,3}
          \and
          Dario Hrupec\inst{4}
          \and
          Dijana Dominis Prester\inst{3}
          \and
          Lovro Pavletić\inst{3}
          \and 
          Saša Mićanović\inst{3}
          }

   \institute{Department of Astrophysics, University of Lodz, Pomorska 149, Łódź, Poland
%   \and 
%North-West University, 11 Hoffman St, Potchefstroom, South Africa
   \and Ru\dj er Bošković Institute, Bijenička cesta 54, Zagreb, Croatia
         \and
         University of Rijeka, Faculty of Physics, Radmile Matejčić 2, Rijeka, Croatia
         \and
         J. J. Strossmayer University of Osijek, Department of Physics, Trg Ljudevita Gaja 6, Osijek, Croatia
             }

   \date{Received XXXX / Accepted YYYY}

% \abstract{}{}{}{}{} 
% 5 {} token are mandatory
 
  \abstract
 % context heading
{The presence of clouds during observations with Imaging Atmospheric Cherenkov Telescopes can strongly affect the performance of the instrument due to additional absorption of light and scattering of light beyond the field of view of the instrument. 
If not corrected for, the presence of clouds leads to increased systematic errors in the results. }
% aims heading
{One approach to correct for the effects of clouds is to include clouds in Monte Carlo simulations to produce models for primary particle classification, energy and direction estimation. However, this method is challenging due to the dynamic nature of cloudy conditions and requires extensive computational resources. The second approach focuses on correcting the data itself for cloud effects, which allows the use of standard simulations. However, existing corrections often prioritise limiting systematic errors without optimising overall performance. By correcting the data already at the image level, it is possible to improve event reconstruction without the need for specialised simulations.}
% methods heading
{We introduce a novel analysis method, based on a geometrical model that can correct the data already at the image level given a vertical transmission profile of a cloud. Using Monte Carlo
simulations of an array of four Large-Sized Telescopes of the Cherenkov Telescope Array, we investigate the effect of the correction on the
image parameters and the performance of the system.
We compare the data correction at the camera level with the use of dedicated simulations for clouds with different transmissions and heights.}
% results heading
{The proposed method efficiently corrects the extinction of light in clouds, eliminating the need for dedicated simulations. Evaluation using Monte Carlo simulations demonstrates improved gamma-ray event reconstruction and overall system performance.}
  % conclusions heading (optional), leave it empty if necessary 
{}

  \iffalse
  \abstract
  % context heading (optional)
  % {} leave it empty if necessary  
   {}
  % aims heading (mandatory)
   {}
  % methods heading (mandatory)
   {}
  % results heading (mandatory)
   {}
  % conclusions heading (optional), leave it empty if necessary 
   {}

\abstract{
Currently used analysis methods for cloud-affected data focus on the correction of the reconstructed energy, however, do not counteract the worsening of the performance parameters. 
We introduce a novel analysis method, based on a geometrical model that can correct the data already at the image level given a vertical transmission profile of a cloud. Using Monte Carlo
simulations of an array of four Large-Sized Telescopes, we investigate the effect of the correction on the
image parameters and the performance of the system.
The proposed method can efficiently correct the effect of light extinction in the cloud without the need to generate dedicated simulations. }
  \fi

   \maketitle

%% \linenumbers

%% main text
\section{Introduction}
\label{sec:intro}

Imaging Atmospheric Cherenkov Telescopes (IACTs) use Cherenkov light induced by cascades of secondary particles developing in the Earth's atmosphere to study very-high-energy (VHE, from 100 GeV to 100 TeV) gamma rays.
However, incorporating the atmosphere into the detector also introduces a susceptibility to fluctuations in transparency and scattering caused by cloud cover.
%However, this makes the atmosphere part of the detector. 
%Thus any changes in the atmosphere's transparency, due to (partial) cloud coverage, will affect this technique. 
Therefore, the presence of clouds during observations can degrade the performance of the telescopes, resulting in significantly increased systematic errors in the measurements if not corrected for.  %Namely, the cloud presence during the observations will degrade the performance of the telescopes and if not corrected, it will lead to increased systematic errors in the measurements.
Depending on the location of the observatory, the atmospheric transmission can also be affected by increased dust concentrations, such as the so-called calima (Saharan dust intrusion, \citealp{2011ACP....11.6663R}). 
However, in this work we concentrate on the case of light loss due to clouds. 

The Cherenkov Telescope Array Observatory (CTAO, \citealp{2013APh....43....3A}) is the next-generation IACT observatory, currently under construction. 
To cover the energy range from a few tens of GeV to hundreds of TeV, CTAO will consist of three types of telescopes:
Large-Sized Telescopes (LSTs), Medium-Sized Telescopes (MSTs), and Small-Sized Telescopes (SSTs).
LSTs are responsible for delivering the best performance in the $\lesssim100$~GeV range \citep{lst1perf}.
The expected performance gain of VHE gamma-ray observations with CTAO, compared to the current generation of Cherenkov telescopes, needs to be matched with analysis techniques to minimise the systematic uncertainties. 
This involves the development of techniques to analyse data taken in the presence of clouds.

Atmospheric conditions can be monitored with the use of a number of devices \citep[see e.g.][]{Fruck14d2, Hahn14f2,2015EPJWC..8902005D,Bergon16c2,Valore18b,2019arXiv190908085E,Gaug19a,Iarlori19b1,Paveletic22d1}.
Using LIDAR devices it is possible to evaluate the transparency of the atmosphere at different heights and, in the case of currently operating observatories, to correct the reconstructed energy and spectral parameters \citep[see e.g.][]{Nolan10f1, Devin19c1,Dorner09g1,Fruck14d2,Fruck15g2, 2022JPhCS2398a2011S, Schmuckermaier23e}.
This information can be used in two different ways.
Firstly, the reduced transmission of the atmosphere can be included in the Monte Carlo (MC) simulations used to generate models to classify the shower's primary parent particle, and to estimate its energy and directions, as well as for the generation of the instrument response functions \citep{2023arXiv230202211P}.
This is the most precise approach and allows the use of the standard analysis chain.
However, this approach is difficult to realise, as the cloud conditions can change even on minutes time scales.
\cite{Schmuckermaier23e} estimate that $<22-44\%$ of cloud observations are hampered by rapid cloud movement. 
The generation of dedicated MC simulations for fine time bins would require huge computer resources and thus is not feasible. 
For example, run-wise simulations applied for five H.E.S.S. telescopes require 300 -- 2000 hours of HEP-SPEC06 CPU time to describe a single 28-minute long run \citep{2020APh...12302491H}. 
In some cases, like real-time analysis (see e.g. \citealp{rtaproc}), it is not possible at all.
Additionally, the analysis chain might become sub-optimal if the images are strongly distorted (e.g. are not parameterised well with a Hillas ellipse,  \citealp{Hillas85}) due to the extinction of the light in the cloud. 
This might be particularly problematic for analysis chains based on semi-analytic shower models \citep[see e.g.][]{2009APh....32..231D}.
Secondly, the data itself are corrected for the effects of the cloud. 
This allows the use of the standard MC simulations. However, the applied correction is often only focused on limiting the systematic errors, without optimising the performance (see e.g. \citealp{Schmuckermaier23e}). 

The main effect of a cloud presence is the dimming of the shower image resulting in a reduced \textit{intensity} parameter and, consequently, an underestimation of the energy. However, other image parameters are affected as well 
\citep{2014JPhG...41l5201S}. 
This can additionally lead to the misclassification of genuine gamma-ray events as hadrons and underestimation of the corresponding flux \citep[see e.g.][]{2020APh...12002450S}.
The correction of the estimated energy needs to take into account the height distribution of the emitted light, which determines what fraction of light is affected by the cloud of a given transmission.
One approach is exploiting the reconstructed height of the shower maximum (from stereoscopic reconstruction), applying the typical longitudinal spread of the shower and computing reconstructed-shower-averaged expected attenuation by the cloud \citep{Schmuckermaier23e}. % \citep{2022JPhCS2398a2011S}.
However, as shown in e.g. \citet{2020APh...12002450S}, the presence of clouds can affect the reconstruction of the height of the shower maximum.
Therefore, \citet{2020APh...12002450S} proposed a different approach, using the average profiles of the Cherenkov emission for a clear atmosphere to derive the effective attenuation and correct the reconstructed energy with an analytical formula that described the expected energy bias. 
Both types of energy bias corrections are limited to high-level analysis -- they are not able to improve e.g. the gamma/hadron separation since they do not improve either the individual image parameterization or the stereoscopic reconstruction of the event geometry.

We propose an alternative method, belonging to the second class of methods, that does not require dedicated MC simulations for every possible cloud transmission profile, while still being able to improve the image parameterization and stereoscopic reconstruction.
It is based on a simple geometrical model relating the pixel position (i.e. the direction in the sky from which the Cherenkov photons are gathered) to the expected height at which Cherenkov photons were emitted.
We perform dedicated MC simulations to study whether the method can recover the image parameters corresponding to observations without clouds.
We also evaluate the expected high-level performance parameters when the method is applied to the data taken in the presence of clouds at different altitudes and with various transparencies.

\section{Monte Carlo simulations}
\label{sec:mcs}

We simulated the development of air showers with CORSIKA 7.7410 \citep{Heck:1998, Bernlohr:2008} induced by gamma rays arriving from an angle of 20 degrees in zenith and 180 degrees in azimuth (counting clockwise from geographic north). The gamma-ray-induced air showers treated with the EGS4 \citep{Nelson:1985} electromagnetic interaction model were simulated at an offset of 0.5 degrees from the centre of the camera. The background protons (treated with the hadronic models UrQMD \citep{Bass:1998ca} and QGSJET-II -04 \citep{Ostapchenko:2004qz} for low-energy and high-energy interactions, respectively) and background electrons (treated with EGS4) were simulated with a cone with a half-opening angle of 10 degrees around the aforementioned zenith and azimuth angles. The simulated energy range for gamma rays and electrons is from 10\,GeV to 100\,TeV and for protons from 20\,GeV to 300\,TeV, both with a spectral index of $-2$.

\begin{figure}[h!]
    \centering
    \includegraphics[width=0.9\columnwidth]{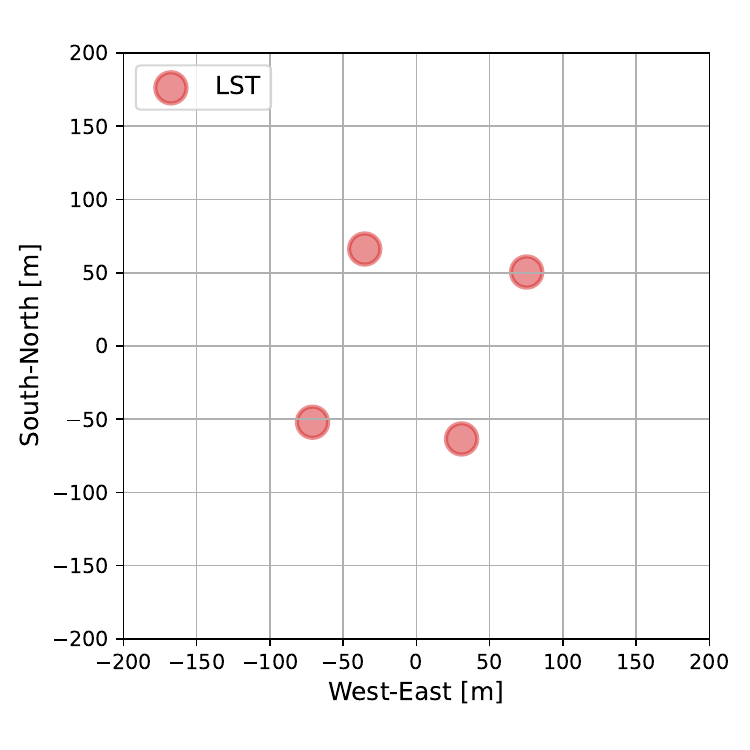}
    \caption{Simulated array of telescopes consisting of 4 Large-Sized Telescopes (LSTs) located on the Canary Island of La Palma at 2156 m above sea level.}
    \label{fig:layout}
\end{figure}

The \texttt{sim\_telarray} \citep{Bernlohr:2008} code 
%with the-so called CTA Prod6\footnote{\url{https://www.mpi-hd.mpg.de/hfm/CTA/MC/Prod6/}} technical description of the telescopes 
%was used to simulate the effects of the atmosphere on the propagation 
was used to simulate the effects of the extinction in the atmosphere  
of the Cherenkov light (emitted in the wavelength range from 270 nm to 750 nm) and the response of the detector (see also \citealp{2000APh....12..255B}). The simulated telescope layout, consisting of 4 LSTs, is shown in Fig. \ref{fig:layout}. For simplicity, we have chosen to simulate all four LSTs with the technical design corresponding to LST-1 (i.e. the operating prototype built on La Palma) of the so-called CTA Prod6 settings. However, this does not impose any additional constraints on the method itself. The maximum impact parameters from the centre of the array and other simulation parameters are summarised in Table~\ref{tab:corsika_params}.

\begin{table}[h!]
\centering
\caption{Parameters used in the input CORSIKA card. Namely, the energy range of the simulated power law and its index, number of simulated showers, half-opening angle of the isotropic simulation (i.e. offset from the centre of the camera for gamma-ray showers), number of reuses of each shower, maximum simulated impact from the centre of the array. Note that the offset of gamma-ray showers from the centre of the camera is achieved by simulating them by setting such VIEWCONE parameter so that the inner and outer radii are nearly identical (i.e., from 0.499~deg to 0.501 deg), effectively defining a specific region (the "ring") in which the showers are generated.}
%\scriptsize
\begin{tabular}{@{}lccc@{}}
\toprule
                   & \textbf{gamma rays} & \textbf{protons} & \textbf{electrons} \\ \midrule
ERANGE (TeV)       & 0.001 -- 100                          & 0.002 -- 300                       & 0.001 -- 100                         \\
ESLOPE     & $-2$                                   & $-2$                                & $-2$                                  \\
NSHOW              & $10^7$                        &  $2\cdot 10^8$                    &  $2\cdot 10^7$                                    \\
VIEWCONE (deg)     & 0.499 - 0.501                                & 0 -- 10                                & 0 -- 10                                  \\
NSCAT              & 10                                   & 20                                & 20                                  \\
CSCAT (m) & 600                                  & 1400                              & 900                                  \\ \bottomrule
\label{tab:corsika_params}
\end{tabular}
\end{table}

We have simulated modified atmospheric profiles based on the U.S. 1976 Standard Atmosphere \citep{1992P&SS...40..553N} using the MODerate resolution atmospheric TRANsmission (MODTRAN) code version 5.2.2 \citep{berk1987modtran,Berk2005MODTRAN5A,stotts2019atmospheric}. MODTRAN is a radiation transfer algorithm that calculates the characteristics of spectral absorption, transmission, emission, and scattering in the atmosphere at moderate spectral resolution (from infrared to ultraviolet). Additionally, users can customise the atmospheric profiles by including their input data. The package also incorporates various parameters that define the aerosol background, including the availability of different aerosol types such as fog, urban, rural, desert, and navy maritime. Furthermore, the properties and species of clouds can be specified in the model. The basic approach is to model the atmosphere as a series of layers of homogeneous density. To obtain the vertical optical depth, the extinction coefficients for each atmospheric constituent are multiplied by the amount of extinction species in each layer. MODTRAN includes the effects of molecular continuum absorption, molecular scattering and absorption, aerosol absorption and scattering. The optical depth of the layer is converted to the transmission through that layer, while the total transmission is determined as the product of the transmission of the individual components and the transmission associated with the scattering attenuation and the continuum \citep{Maghrabi:2007}. %Finally, the total radiance is calculated using the radiative transfer equation. 

According to \cite{Fruck:2022}, 92\% of the cloud covers over La Palma during observations with MAGIC telescopes are single-layer clouds, with the lower, optically-dense clouds being predominantly cirrus, cumulonimbus, and altostratus. 
%In spring and winter, bases of clouds over La Palma are usually located between 8 and 10 km a.s.l., corresponding to $\approx$ 5.8 and $\approx$ 7.8 km a.g.l., respectively, and show a multimodal distribution of vertical optical depth with peak values at 0.09 and 0.5 (transmissions of $\approx$ 90\% and $\approx$ 60\%, respectively, all up to $\approx$ 30\%). 
In spring and winter, bases of clouds over La Palma are usually located between $\approx$ 6 and 8 km above ground level (a.g.l.).
They show a multimodal distribution of vertical optical depth with peak values at 0.09 and 0.5 (transmissions of $\approx$ 0.9 and $\approx$ 0.6, respectively). 
The bases of the summer clouds are concentrated around 6 km a.g.l., with a similar distribution of vertical optical depth. 
The typical geometrical thickness of the cloud ranges from 1 to 4 km. 
Based on the available information from the aforementioned study, we simulated several independent atmospheric models, including single-layer clouds with a total transmission of 0.4, 0.6, or 0.8. For a common transmission of clouds on La Palma ($T=0.6$), we simulated the cloud base at 5, 7, or 9 km a.g.l. The cloud with transmission of 0.4 or 0.8 was simulated only at 7 km a.g.l. since this is the most common height of clouds on La Palma. 
As a baseline model, the cloudless atmospheric profile ($T=1$) was also simulated.
The details are summarised in Table~\ref{tab:modtran}. High layers of clouds are relatively thin compared to the longitudinal spread of air showers (several kilometres), so we can assume them to be homogeneous without considering their internal structure. Therefore, we decided to model grey (wavelength-independent) altostratus clouds with a thickness of 1 km. This approach allowed better control over parameters that could potentially affect the telescope performance. 
To evaluate the performance of the method in a more complicated case we also simulated a case with two clouds at different heights.
In this case, we assumed that the first layer occurs from 5 to 6 km a.g.l, and the second from 7 to 8 km a.g.l. 
Each of the two layers has a transmission of 0.8, resulting in the total transmission through the two clouds of 0.64. 

\begin{table}[h!]
\centering
\caption{Parameters of simulated clouds: transmission, the height of the base (above ground level) and geometrical thickness.
The first line corresponds to a cloudless condition, while the last one to a two-layer cloud. }
%\scriptsize
\begin{tabular}{@{}ccc@{}}
\toprule
Transmission & Base Height (km) & Thickness (km)\\ \midrule
1.0          & -    & -       \\
0.8          & 7    & 1           \\ 
0.6          & 5    & 1     \\
0.6          & 7    & 1     \\
0.6          & 9    & 1     \\
0.4          & 7    & 1           \\
0.8 \& 0.8 & 5 \& 7 & 1 \& 1 \\
\bottomrule
\end{tabular}
\label{tab:modtran}
\end{table}

\section{Image correction model}\label{sec:model}
An image of an event registered by an IACT reflects the propagation of the shower through the atmosphere.
%Due to the change of the refractive index with the height, the longitudinal profile of the shower is encoded along the main axis of the image (the so-called Hillas ellipse). 
The change of the refractive index of the atmosphere with the height reflects in the height dependence of the Cherenkov angle. 
Therefore, the longitudinal profile of the shower is encoded along the main axis of the image (the so-called Hillas ellipse). 
The ``head'' of the shower image is composed mostly of light emitted in the top parts of the atmosphere, while the ``tail'' reflects the emission produced closer to the telescope. 
Therefore, the observed Cherenkov image of the shower is not homogeneously dimmed by the cloud, but rather the ``head'' part of the shower is more strongly affected, influencing the resulting shape (and derived Hillas parameters) of the image \citep[e.g.][]{2014JPhG...41l5201S}. 

\begin{figure}[t]
    \centering
    \includegraphics{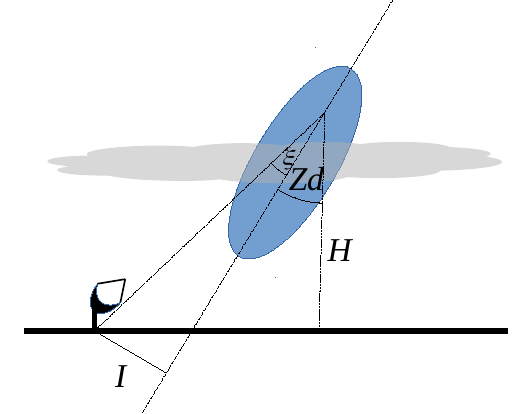}
    \caption{Sketch of the assumed geometry. The blue ellipse represents the shower and the dashed line is its longitudinal axis. For the impact parameter, $I$, of the shower with zenith angle distance, $Zd$, the Cherenkov photons emitted at the height, $H$, will be registered by the telescope at an angular distance, $\xi$, from the primary particle direction.}
    \label{fig:geometry}
\end{figure}

To correct the data for the effects introduced by Cherenkov light extinction in the cloud, we developed a simple geometrical model (see Fig.~\ref{fig:geometry}).
Using basic shower geometry parameters, the model relates the pixel position on the camera (corresponding to a particular direction in the sky from which Cherenkov photons are observed) to the expected height of the emitted Cherenkov light that is registered by that pixel 
%(assuming $I \ll H$)
(assuming $I \sin Zd \ll H$):
\begin{equation}
    \xi = \arctan\bigg(\frac{I}{H} \cos{Zd}\bigg).\label{eq1}
\end{equation}
$\xi$ is the offset angle from the primary gamma-ray direction corresponding to the height $H$ (measured a.g.l.), $Zd$ is the zenith angle of the observations (a proxy for the zenith angle of the shower) and $I$ is the preliminary reconstruction of the impact parameter.
A similar method of relating the position in the camera to the longitudinal distribution of the shower (but applied in the context of cosmic ray studies) was used by \citet{2023APh...14802817G}.
Such an approach allows us to apply the pixel-by-pixel correction already at the image level, and propagate the corrected image through a standard IACT analysis chain.

\begin{figure*}[t!]
    \centering
    \includegraphics[width=0.99\textwidth]{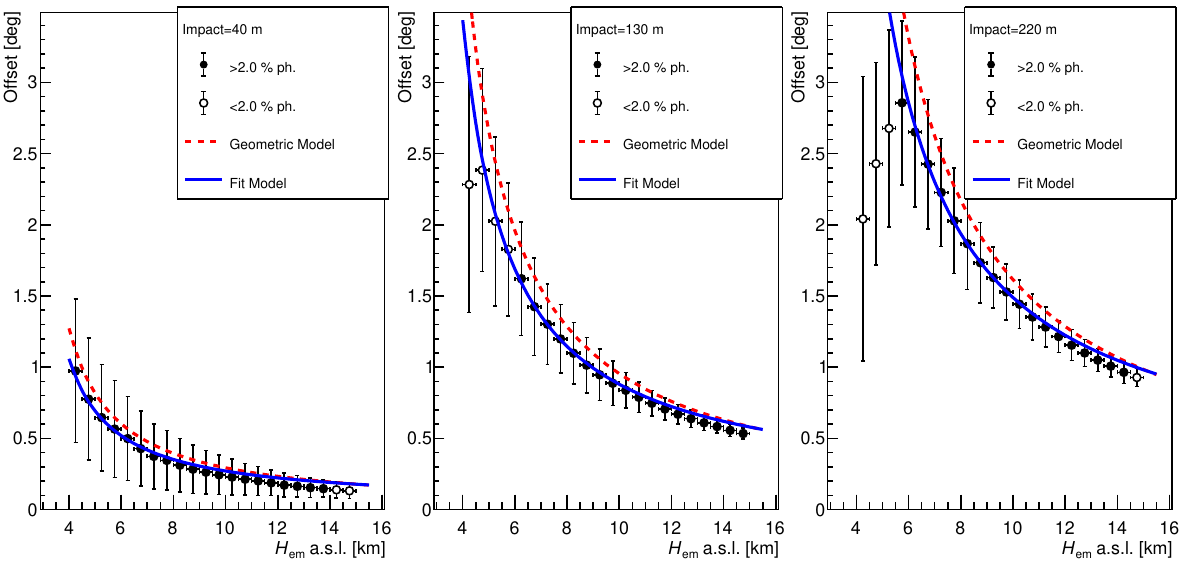}
    \caption{ Offset angle of the photons hitting the observation level from different emission heights ($H_{\textrm{em}}$, measured above sea level) for vertical 100~GeV gamma rays.
    Points show the average (markers) value of the offset measured along the true main axis of the shower, and the corresponding standard deviation (error bars).
    Empty markers report the heights that do not contribute considerably to the total image (below 2\% of the total). 
    The pure geometrical model described by Eq.~\ref{eq1} is shown as the red dashed curve, while the global fit model (Eq.~\ref{eq2}) is expressed as the solid blue line. 
    The left, middle and right panels show the results for the impact parameters of 40~m, 130~m and 220~m respectively.
    }
    \label{fig:model_offset}
\end{figure*}

To validate and eventually improve the geometrical model, we have performed dedicated CORSIKA simulations of vertical gamma rays observed at different energies and impact parameters. 
To generalise the study, these simulations are not applying any telescope-specific effects, such as photoelectron conversion probability, or the optical point-spread function. 
We have applied Rayleigh and Mie scattering in the atmosphere according to a formula given in \citet{1989iuec.book.....S} and computed the average offset angle of the Cherenkov photons from different ranges of emission height. 
An example of these calculations is shown in Fig.~\ref{fig:model_offset}.
While the model defined by Eq.~\ref{eq1} describes the shape of the dependence of the average offset $\xi$ of photons emitted at a given height $H$, it slightly overestimates it by $\sim10-20\%$.
This is likely related to the lateral and angular distribution of charged particles that emit Cherenkov photons, which is ignored by the geometrical model.
To improve the model, we correct this minor bias by introducing a phenomenological correction factor:
\begin{equation}
    \xi'= \xi\left(0.877 + 0.015 \frac{H + H_0}{7\,\mathrm{km}}\right), \label{eq2}
\end{equation}
where $H_0=2.2$~km is the height of the telescopes above sea level, and $\xi'$ is the corrected offset angle.
Notably, the bias correction factor (the parentheses part of Eq.~\ref{eq2}) does not depend on the energy of the primary gamma ray, nor on the impact parameter, but describes the average offset angle very accurately (the differences are mostly within a few \%). 
Nevertheless, it was optimised for low-zenith angle observations, and it might require modifications for observations at higher zenith distance angles.
For large impact values, a larger deviation from the model curve is present at very low emission heights. 
This is an artefact of clipping the calculations at the angle of $3.5^\circ$ from the primary particle direction.
However, photons with such high offsets would not be registered by LST telescopes, due to the size of their field of view.
Additionally, the emission from such low altitudes for high impact values contributes only a small fraction of the total light yield from the shower. 

To derive the average offset from Eq.~\ref{eq2} and to associate it to particular pixels in the camera, the tentative geometry of the shower (namely the first estimation of the arrival direction and the impact parameters with respect to all telescopes) needs to be used. 
It is obtained from the uncorrected image using the classical axis crossing method \citep{1999APh....12..135H}. 
Due to the symmetry, cloud attenuation will not introduce large differences in the attenuation perpendicular to the main axis of the image. 
Therefore, the orientation of the main axis of the image is not expected to be biased by the cloud attenuation, apart from the increased statistical uncertainty due to the smaller number of registered photons. Methods exploiting the image shape (e.g. the DISP method, \citealp{2001APh....15....1L}) would suffer additional systematic errors due to the effect of the cloud on the \textit{Length} parameter, which would bias the reconstructed source position. 
Therefore, the classical axis crossing method would be more robust against cloud extinction to obtain the preliminary stereoscopic parameters. 

We associate the corresponding emission height to each pixel by computing its centre's projected distance from the tentative primary gamma-ray direction, computed along the main axis of the image. 
Next, we exploit the knowledge of vertical transmission $T$ from each height, which could be easily measured during real observations typically with a LIDAR instrument even every few minutes. 
Since the geometrical path through the cloud is increased due to the zenith distance angle $Zd$, the actual transmission affecting the Cherenkov light is $T^{(1/\cos Zd)}$.
%In a case of a narrow cloud at a height $H_c$ and the total transmission $T_c$, this results in dividing the camera in two parts, perpendicularly to the main shower axis with the crossing line at the distance of $\xi'$ for $H=H_{c}$.
%The ``head'' part of the image is then up-scaled by $1/T_c$, while the ``tail'' part is left unchanged. 
Therefore, for each event, the camera is divided into stripes, perpendicular to the main shower axis, each corresponding to a particular height. 
%The ``head'' part of the image (corresponding to heights above the top of the cloud) is then up-scaled by $1/T_c^{1/\cos Zd}$, while the ``tail'' part (height below the bottom part of the cloud) is left unchanged, with an intermediate correction in the partially affected heights in the middle of the cloud. 
The part of the image corresponding to heights above the top of the cloud is then up-scaled by $1/T_c^{1/\cos Zd}$ (where $T_c$ is the total transmission of the cloud), while the part corresponding to the height below the bottom part of the cloud is left unchanged, with an intermediate correction in the partially affected heights in the middle of the cloud. 
This way, the model easily accommodates any transmission profile (including extended or multiple clouds).
In the case of very low clouds, or e.g. calima,  where all photons are emitted above the cloud, the model simply up-scales the entire image by the same factor.
%In this case we split the camera in stripes perpendicular to the main axis of the image. 
%Whole stripe, will be then corrected with the same transmission value (at the corresponding height).

The corrected images are cleaned 
%(with the same or updated cleaning mask) 
and parameterised. Finally, further stages of the analysis, i.e. the final reconstruction of the arrival direction, the energy estimation and the classification of gamma/hadron events, can be performed.

%\komm{... description of the geometrical model with formulae, phenomenological correction, comparison with simple MCs...}

%\begin{figure}
%    \centering
%\vspace{5cm}
%%    \includegraphics{}
%    \caption{Comparison of the angular distance between the primary gamma ray direction and the Cherenkov photons emitted at different heights compared with the geometrical model.}
%    \label{fig:mc_model}
%\end{figure}

\subsection*{Implementation in LST analysis chain}

The generated MC simulations of all primary particles and both atmospheric conditions (cloudless and clouds) were treated in the same way and were analysed with the \texttt{ctapipe}\footnote{\url{https://github.com/cta-observatory/ctapipe}} version 0.12 and \texttt{lstchain}\footnote{\url{https://github.com/cta-observatory/cta-lstchain}} version 0.9.13, constituting a prototype of the low-level CTAO data processing approach \citep{ctapipe, lstchain}. First, we reduced the data from the raw data (R0) to the data level 1 (DL1) using the \texttt{r0\_to\_dl1} script modified to calculate stereo parameters.
%\footnote{\url{https://github.com/cta-observatory/cta-lstchain/blob/master/lstchain/reco/r0\_to\_dl1.py}}. 
In this step, we applied the camera calibration, extracted the images and cleaned them to distinguish the pixels dominated by the Cherenkov light from the background by applying the tailcut method with the standard parameters, i.e. \texttt{picture threshold} = 8, \texttt{boundary threshold} = 4 and a minimum number of pixel neighbours of 2. Subsequently, the shape, size and orientation of each given shower image were parameterised by the Hillas parameters.
After this step, we obtained sets of the parameters describing the first and second moments of the charge distribution as well as e.g. timing or truncation for each registered image. 
%This allowed us to estimate the distance between the center of the ellipse in the camera and the putative shower arrival direction using the disp method \citep{2001APh....15....1L}. 
Moreover, we derived the indicative stereo reconstruction parameters based on the previously calculated moments, employing the \texttt{HillasReconstructor} class implemented in \texttt{ctapipe}. Having the extracted images as well as the DL1 and stereo parameters of the events, the image correction model was applied to the data (following the prescription described in Section~\ref{sec:model}) with the clouds.
Finally, the corrected images are cleaned, followed by the Hillas parameterization and the calculation of the stereo parameters.

To reconstruct the primary particle energy, its arrival direction, and separate the particle types, i.e. very-high-energy photons from hadrons, the random forest (RF) method was adopted. The RF models were trained using simulated gamma rays and protons, using \texttt{build\_models} function implemented in the \texttt{dl1\_to\_dl2} script. 
%with around $8.5 \cdot 10^6$ gamma-initiated  events, 
%We trained the RFs using four estimators, namely energy and disp regressors.
For the first two tasks we trained RF regressors for the reconstruction of the energy and the distance between the shower image centre and source position, the so-called \textit{disp} parameter. 
For the third task, the RF classifier was used.
%In order to train parent particle classifiers the additional simulation sample consisting of $2.8 \cdot 10^6$ proton-initiated events was used as well. 
%via the \texttt{dl1\_to\_dl2} script\footnote{\url{https://github.com/cta-observatory/cta-lstchain/blob/master/lstchain/reco/dl1_to_dl2.py}}. 
Next, with \texttt{apply\_models} function in the aforementioned \texttt{dl1\_to\_dl2} script, we applied the trained RF models to the independent samples of MC simulations of cloudless sky, of different clouds, and the corrected ones, obtaining the DL2 reconstructed parameters for each set of data.

Depending on the method applied, the RF models were calculated based on the transmission $T = 1$ data only (in case of lack of correction or the correction applied to images themselves) or on corresponding cloud data (if the correction is applied by using dedicated MC simulations, like in \citealp{2023arXiv230202211P}).

\begin{figure*}[t]
\includegraphics[trim=100 50 100 50, clip,width=0.99\textwidth]{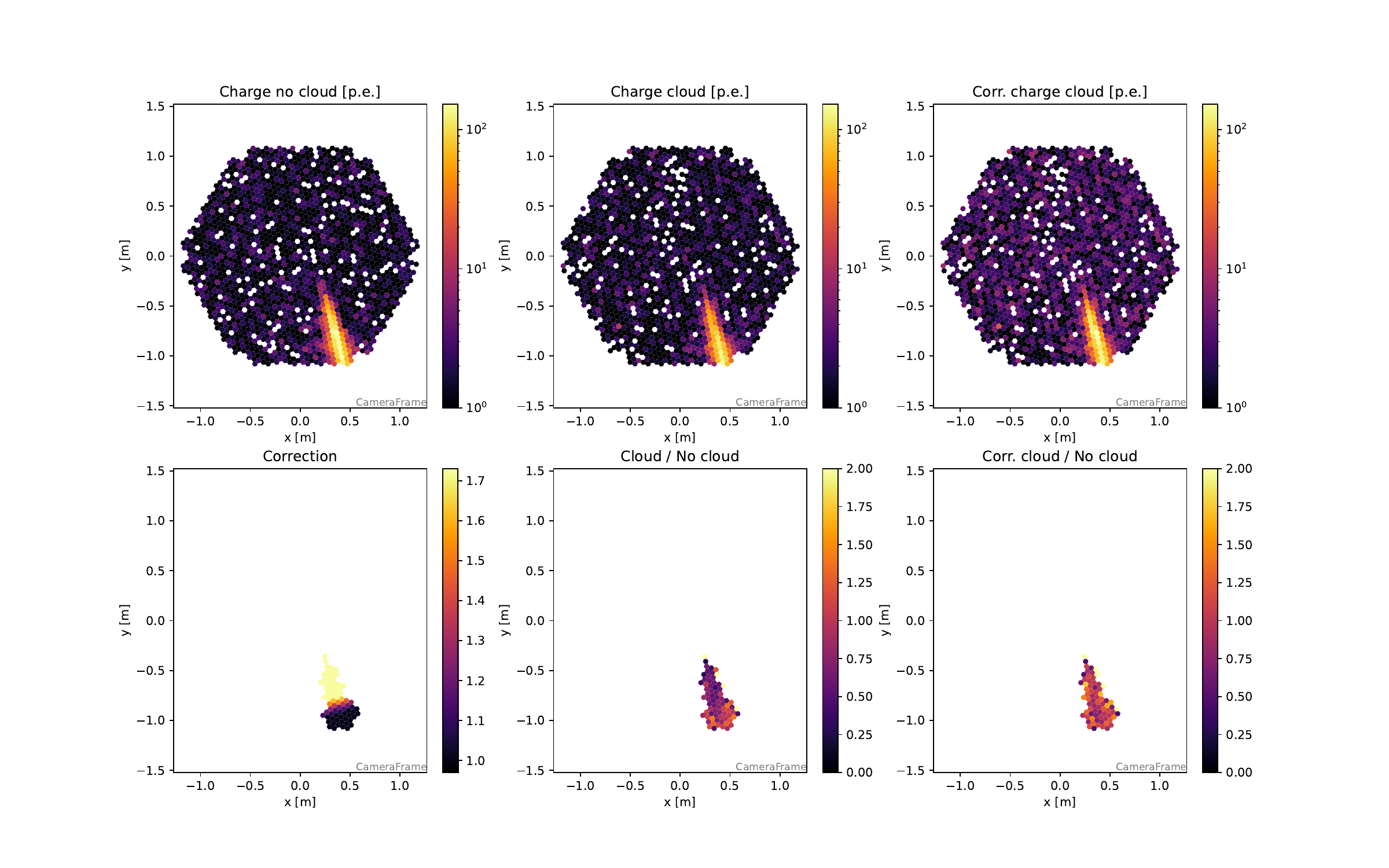}
    \caption{Example of the image correction for the attenuation of light in a cloud. 
    A cloud with height of the base at 7 km a.g.l., thickness of 1 km and transmission of 0.6 is simulated. 
    The top-left panel shows the shower image with cloudless observations, the top-middle panel shows the same shower simulated with additional light attenuation in the cloud, and the top-right shows the corrected image.
    The bottom-left panel shows the correction factor applied, the bottom-middle panel shows the ratio of reconstructed signals of uncorrected image to cloudless image and the bottom-right panel shows the same between corrected and cloudless images. 
    For clarity, only pixels surviving the cleaning for the case of observation with the cloud are shown in the bottom panels.
    ``Head'' of the shower is in the top part of its image.  
    }\label{fig:event}
\end{figure*}

An example event image corrected with this method is shown in Fig.~\ref{fig:event}.
As expected (see the bottom middle plot), the simulations with the cloud result in reduced light yield in the ``head'' part of the shower.
Since the simulations are done independently (including the randomisation of the night sky background, NSB, and individual photoelectron registration), there are also small statistical differences in the ``tail'' part of the image but on average this part is not attenuated. 
After the correction, the ``head'' part of the image recovers the light yield level of the no-cloud simulations. 
Notably, the correction is also applied to pixels dominated by the NSB noise in the part of the camera corresponding to heights above the cloud (compare the top middle and top right panels). 
This increases the noise level of the image. 
Therefore, with a large transmission correction, the original cleaning thresholds might not be sufficient to suppress the pixels with signals dominated by the NSB signals.
We test two different approaches to this problem.
In the first approach, we maintain the original cleaning mask (obtained from the uncorrected cloud-affected image). 
While this approach is intended to prevent noise-dominated pixels from remaining as part of the image, it can also exclude some pixels whose charge is just below the threshold due to light attenuation.
In the second approach, we redo the cleaning with increased picture and boundary thresholds by a factor of $1/T_c$ (we refer to this approach as ``additional cleaning''). 
To ensure the self-consistency of the analysis, the same cleaning needs to be applied to the data (simulated in this work) as for the MC simulations used for RF-based training or calculation of the instrument response functions.  
The second approach is expected to improve the match between image parameters derived from uncorrected simulations without clouds and corrected simulations with clouds, however, at the price of increased analysis threshold, in particular for low-transmission clouds. 
%In the next section, we evaluate both methods.
%We also compare them to analysis based on cloud-affected Monte Carlo simulations, to estimate both their performance and the induced systematic errors. 

%%%%%%%%%%%%
%%%%%%% MP: I checked and corrected text until the results (on Sept 9, 2023)
%%%%%%%%%%%%

\section{Results}

%\komm{.. describe the different comparisons: no cloud vs cloud without correction (and cloudless RFs) vs cloud with correction vs cloud with correction and redone cleaning vs dedicated analysis with cloud MCs used for RFS...}

To evaluate the performance of the correction with the developed model, we perform different kinds of comparisons.
We compare first the effect on the individual image parameters, and how well they are corrected by the model.
Next, we evaluate the typical performance parameters of the Cherenkov telescopes (sensitivity, angular and energy resolution, and bias in energy estimation).
We compare them between different analysis schemes for cloud-affected data (no correction, data correction with/without additional cleaning, dedicated MC simulations) and for cloudless data. 

\subsection{Image and stereoscopic parameters comparison}
\label{sec:hillas}

\begin{figure*}[tp]
    \centering
    \includegraphics[width=\textwidth]%, height=23 truecm]
    {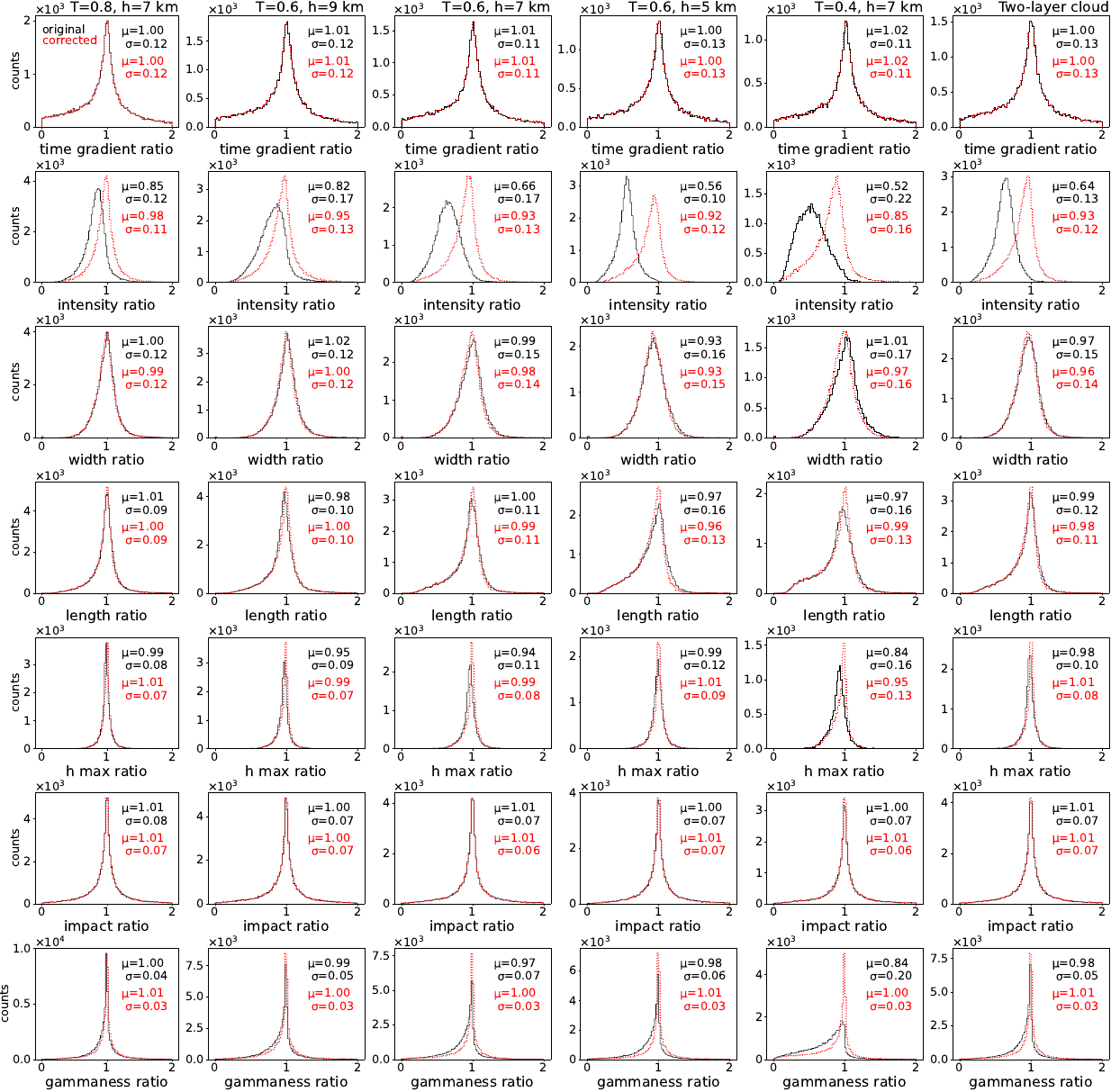}
    \caption{Comparison of ratios of the image and stereoscopic parameters simulated with cloud presence (without correction in solid black lines; with correction, but without redoing the cleaning in red dotted lines) to the values of the same parameters simulated in the cloudless sky.
    Different rows show different parameters. From top to bottom: 
    time gradient measured along the main axis of the image, 
    \textit{Intensity}, 
    \textit{Width},
    \textit{Length},
    height of the shower maximum,
    impact parameter,
    \textit{gammaness}.
    Different columns report different cloud heights a.g.l. and transmissions, from left to right: 
    $T=0.8$, $h=7$~km;
    $T=0.6$, $h=9$~km;
    $T=0.6$, $h=7$~km;
    $T=0.6$, $h=5$~km;
    $T=0.4$, $h=7$~km; and
    two-layer cloud.
    In each panel, the bias and spread are reported as the mean, $\mu$, and standard deviation, $\sigma$, of a Gaussian fitted to the distribution of the ratios.
    }
    \label{fig:ratios}
\end{figure*}

In the first step, we evaluate the impact of the cloud on the selected image parameters that are used in the further stages of the analysis (see also \citealp{2014JPhG...41l5201S}).
In Fig.~\ref{fig:ratios} we show the ratios of these parameters obtained with the same simulated showers observed with the cloud to the cloudless observations for various cloud transmissions and heights in the case of the analysis without redoing the cleaning. 
In the same figure, we also present the corresponding ratios of the corrected image parameters to their cloudless counterparts.
Note that while the comparisons are made with the same simulated showers, the simulations of the telescope response to each Cherenkov photon are governed by random numbers that determine the probability of photon reflection, conversion to p.e., etc. 
Therefore, the parameter ratios are subject to irreducible statistical uncertainties connected with those fluctuations, that would be present even in the case of a perfect correction. 

The parameter most directly affected is the \textit{intensity}, which shows a clear bias (see the black curves) due to the light extinction in the cloud.
If most of the Cherenkov light is emitted above the cloud, the expected drop of intensity is $\sim 1/T_c$ (for observations at low-zenith distance angle). 
This drop may be further enhanced by signals in pixels in the outer parts of the image that fall below the image cleaning threshold. 
The bias introduced by the cloud on the \textit{intensity} parameter is nearly perfectly corrected by the proposed method. 
For example for a cloud with transmission $T=0.6$ at the height of $h=7$~km a.g.l., the bias of $-0.34$ is reduced to only $-0.07$.  Additionally, the spread of the ratio distribution is also decreased (from 0.17 to 0.13 in the same example), showing that the method can correct efficiently images in which different fraction of the signal is affected by the cloud.
For very opaque clouds where the effect is the strongest, the correction decreases the bias considerably but does not eliminate it (e.g. for $T=0.4$ clouds at the height of $h=7$~km a.g.l. the bias is reduced from $-0.48$ to $-0.15$). 

\textit{Width}, \textit{Length} and \textit{time gradient} do not show any large bias (except of the $T=0.6$, $h=5$~km a.g.l. cloud the bias is $\lesssim$3\%). 
%, however the distribution shows some broadness.
%This can be partially due to the fact that the cloudless and cloud simulations are done independently (including the randomness of the noise and p.e. detection). 
Nevertheless, particularly in the case of the \textit{Length} parameter, the distribution of the ratio of corrected to the cloudless case is slightly more peaked.
We interpret this, similarly to the \textit{Intensity} case, as improvement due to corrections of events with a different fraction of image affected by the cloud.
The strongest case of such an effect is seen on $T=0.6$, $h=5$~km a.g.l. and $T=0.4$, $h=7$~km a.g.l. clouds, where the spread is decreased from 0.16 to 0.13. 
Smaller effects are also expected for the \textit{Width} and \textit{time gradient} parameters. 
The former describes the lateral development of the shower. Here, the effect is small, since extinction in a cloud would affect mostly the observed longitudinal distribution of light. 
The \textit{time gradient} parameter describes the evolution of observed arrival time (which is basically not affected by the attenuation in the cloud) along the main axis of the image, hence any effect on this parameter would be of second order. 

Intriguingly, a bias is evident in the estimation of the height of the shower maximum. 
For a $T=0.4$, $h=7$~km a.g.l. cloud it reaches $-0.16$. 
This is expected, as the extinction of the part of the light above the cloud would bias the image centroid outwards from the true direction. 
Therefore, the height of the shower maximum is underestimated if no correction for the presence of the cloud is applied.
This is particularly important, as this parameter is commonly used to evaluate on an event-by-event basis the fraction of lost light due to the cloud. 
The here proposed method successfully corrects the bulk of the bias (even for the most opaque considered cloud, where it is reduced to $-0.05$), and additionally makes the distribution of the ratio to the cloudless case more peaked. 

Finally, the combined \textit{gammaness} parameter, describing how likely the event is a gamma ray is also affected.
In most of the considered cases the spread of the \textit{gammaness} parameter induced by the cloud is 0.04 -- 0.07. 
With the presence of a cloud, it develops a tail towards lower values, which is expected, because the showers look less similar to their cloudless counterparts used in RF training. 
The effect is more pronounced for more opaque clouds, particularly if they are at a height comparable to the typical height of the shower maximum.
In particular for $T=0.4$, $h=7$~km a.g.l. the spread induced by the cloud is 0.2, with an additional bias of $-0.16$. 
The proposed correction method successfully sharpens the obtained distribution of the ratio to the cloudless case. 
In all investigated cases the spread drops down to just $0.03$ with a bias $<1\%$.

%The proposed correction procedure (see green curves) is successful in removing the bias from the intensity and height of the shower maximum.
%Also in other image parameters, the ratio of the corrected to cloudless value of the parameter is more peaked around 1.
%Notably such an improvement is large in the case of the gammaness parameter. 

\begin{figure*}[tp]
    \centering
    \includegraphics[width=\textwidth, height=\textheight, keepaspectratio]
    %, height=23 truecm]\
    {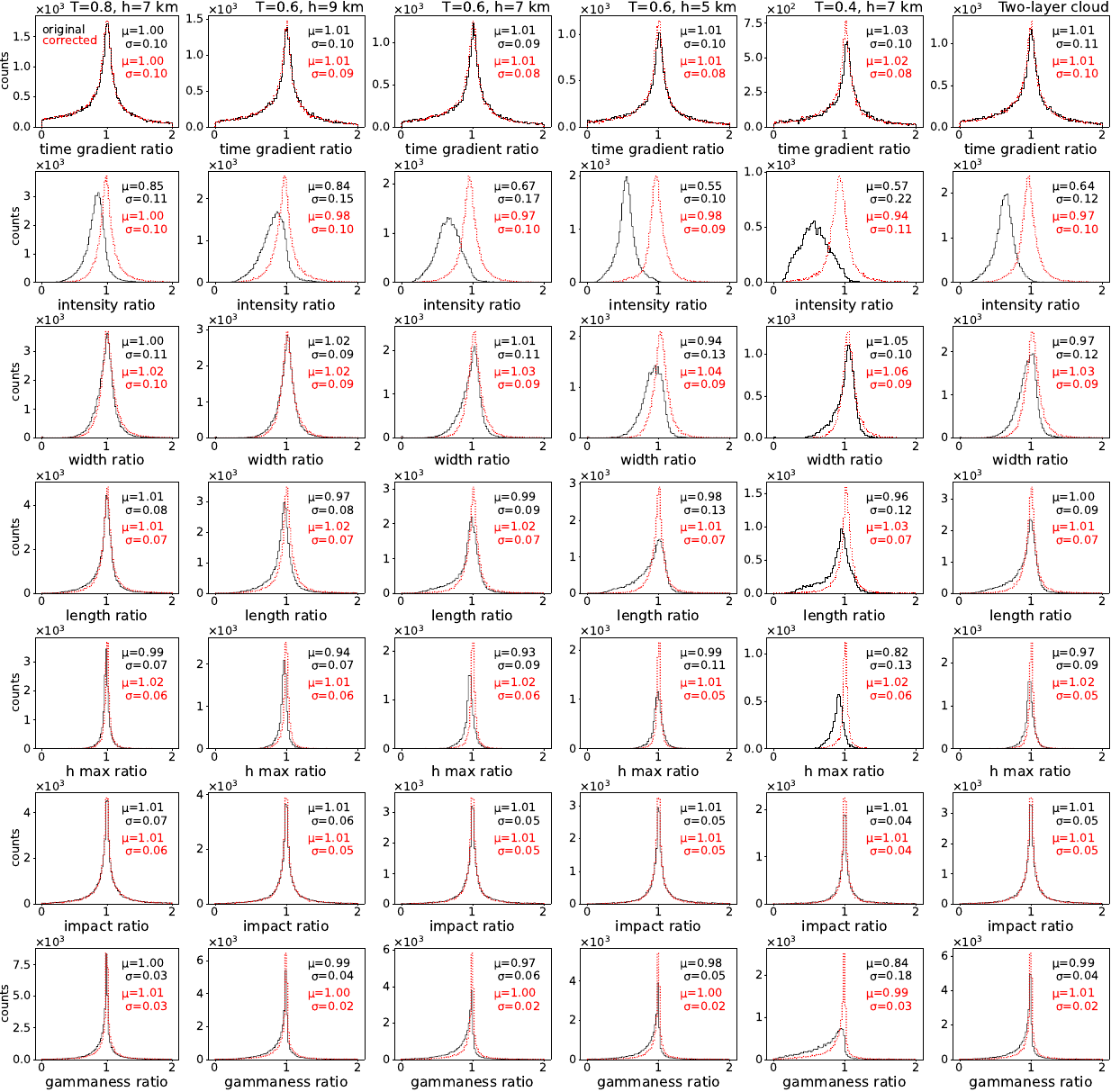}
    \caption{
    Like in Fig.~\ref{fig:ratios}, but reapplying higher cleaning thresholds after the image cloud correction procedure. 
    }
    \label{fig:ratios_clean}
\end{figure*}

In Fig.~\ref{fig:ratios_clean} we show a similar comparison, however applying higher cleaning thresholds, which allow us to redo the cleaning with the corrected image. 
Such an analysis scheme provides a nearly perfect correction of the \textit{intensity} parameter. 
Interestingly, the correction of the image parameters \textit{Width} and \textit{Length} is even better than in the case of lower cleaning thresholds. 
The resulting spread induced by the cloud is as low as 0.07 -- 0.10. 
Similarly, also the correction of the stereoscopic parameters and \textit{gammaness} is further improved with respect to the method without redoing the cleaning.

\subsection{Effective collection area}
\label{sec:coll}

To evaluate the energy-dependent systematic errors introduced by the proposed method, we compute and compare the collection areas (as functions of the true energy of gamma rays) for different types of analysis. 
In the calculations, we apply typical cuts that are used in IACT analysis for gamma/hadron separation.
Namely, we apply a \textit{gammaness} cut that results in 90\% gamma-ray efficiency in each estimated energy bin.
Similarly, the efficiency of the angular cut ($\theta^2$) from the nominal source position is set to 70\%. All performance parameters are obtained with images having \textit{intensity} parameters above  $50$ p.e. 
%The performance is obtained with \komm{???... XXX cuts ???}.

\begin{figure*}[tp!]
    \centering
    \includegraphics[width=0.33\textwidth]{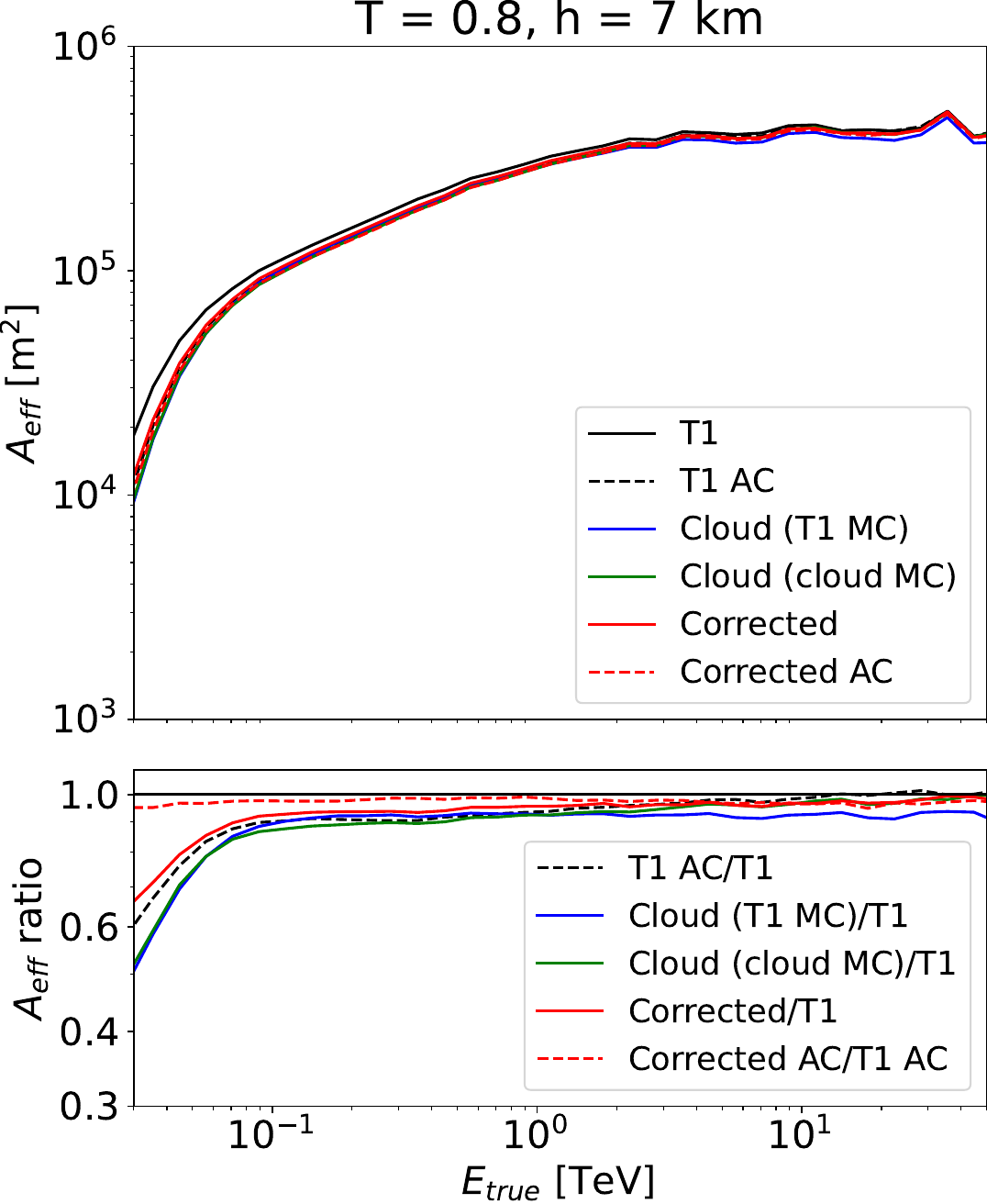}
    \includegraphics[width=0.33\textwidth]{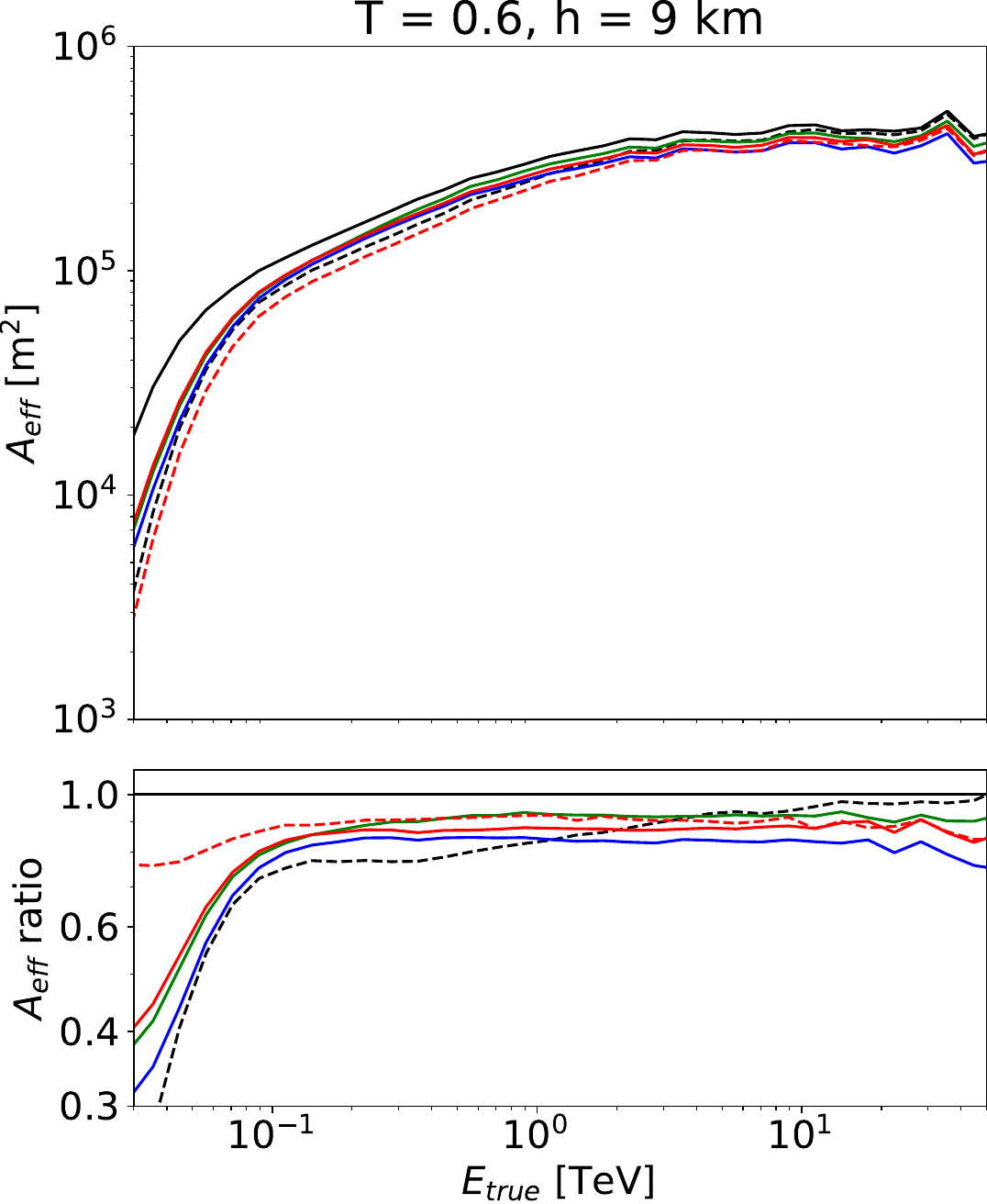}
    \includegraphics[width=0.33\textwidth]{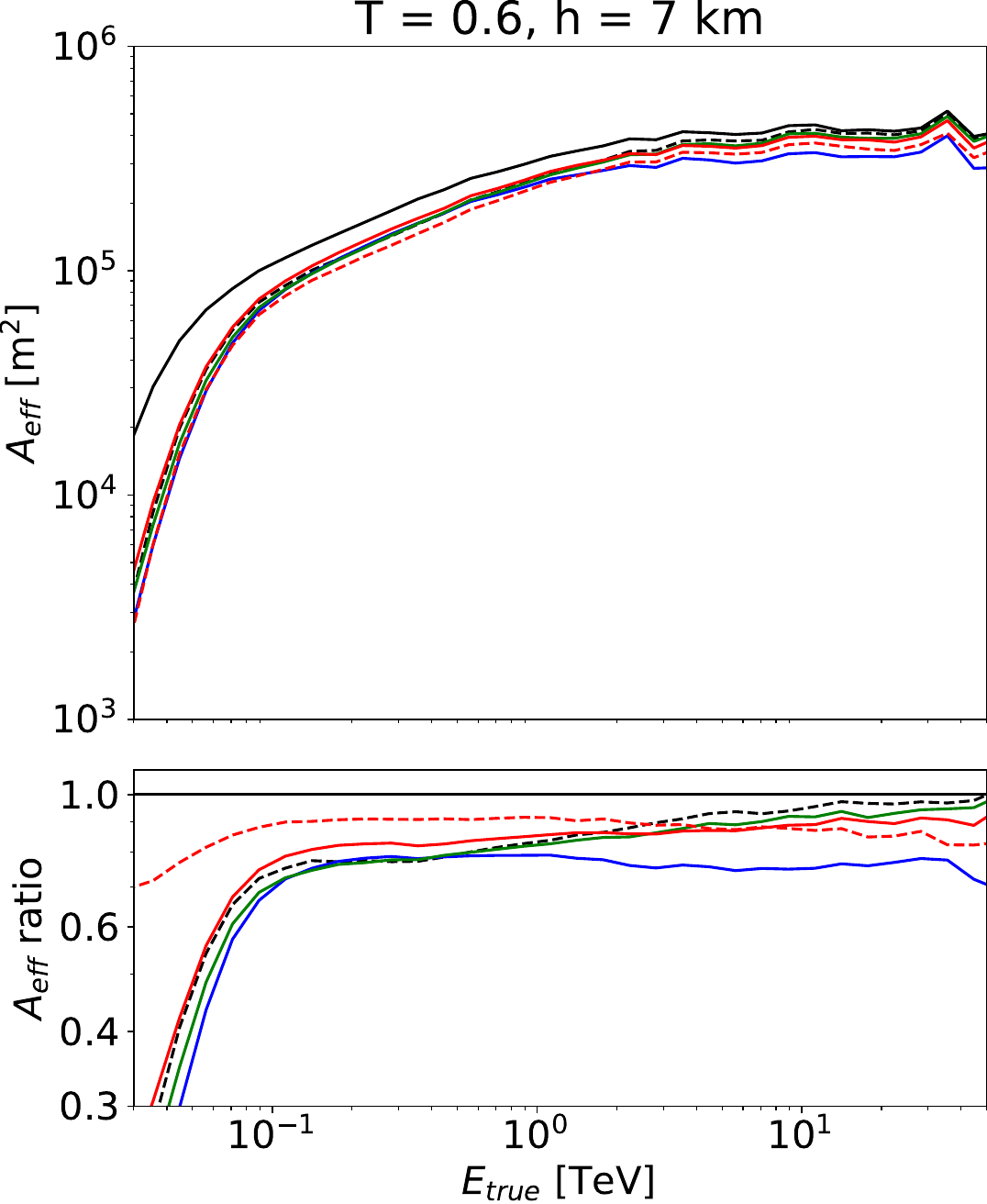}

    \vspace{0.2cm}
    %\includegraphics[width=0.33\textwidth]{figs/Effective_area_T0.8_h7km_geff0.9_theff0.7_ratio.pdf}
    %\includegraphics[width=0.33\textwidth]{figs/Effective_area_T0.6_h9km_geff0.9_theff0.7_ratio.pdf}
    %\includegraphics[width=0.33\textwidth]{figs/Effective_area_T0.6_h7km_geff0.9_theff0.7_ratio.pdf}
    %\hline
    %\rule{\textwidth}{0.5 pt}
    \includegraphics[width=0.33\textwidth]{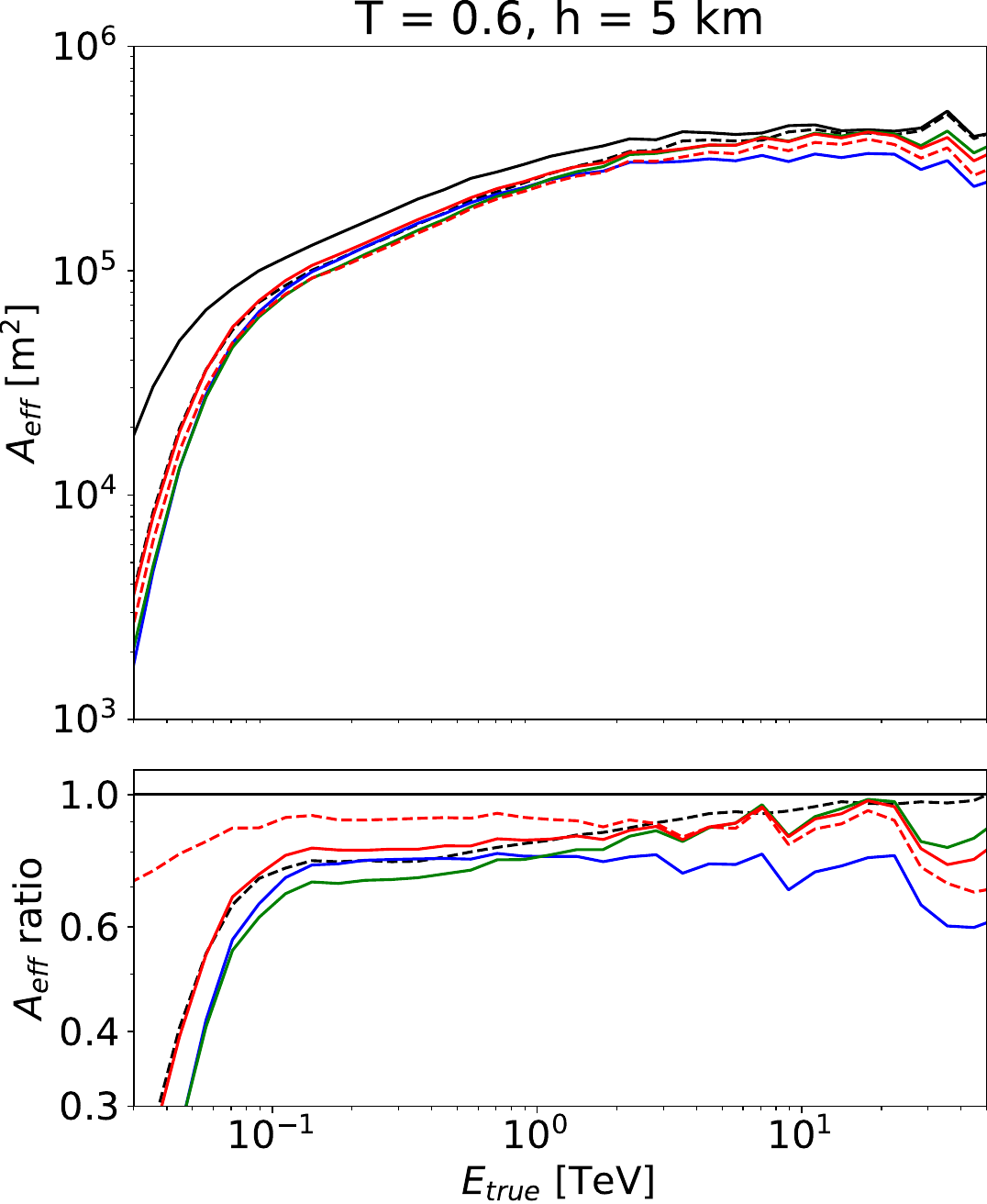}
    \includegraphics[width=0.33\textwidth]{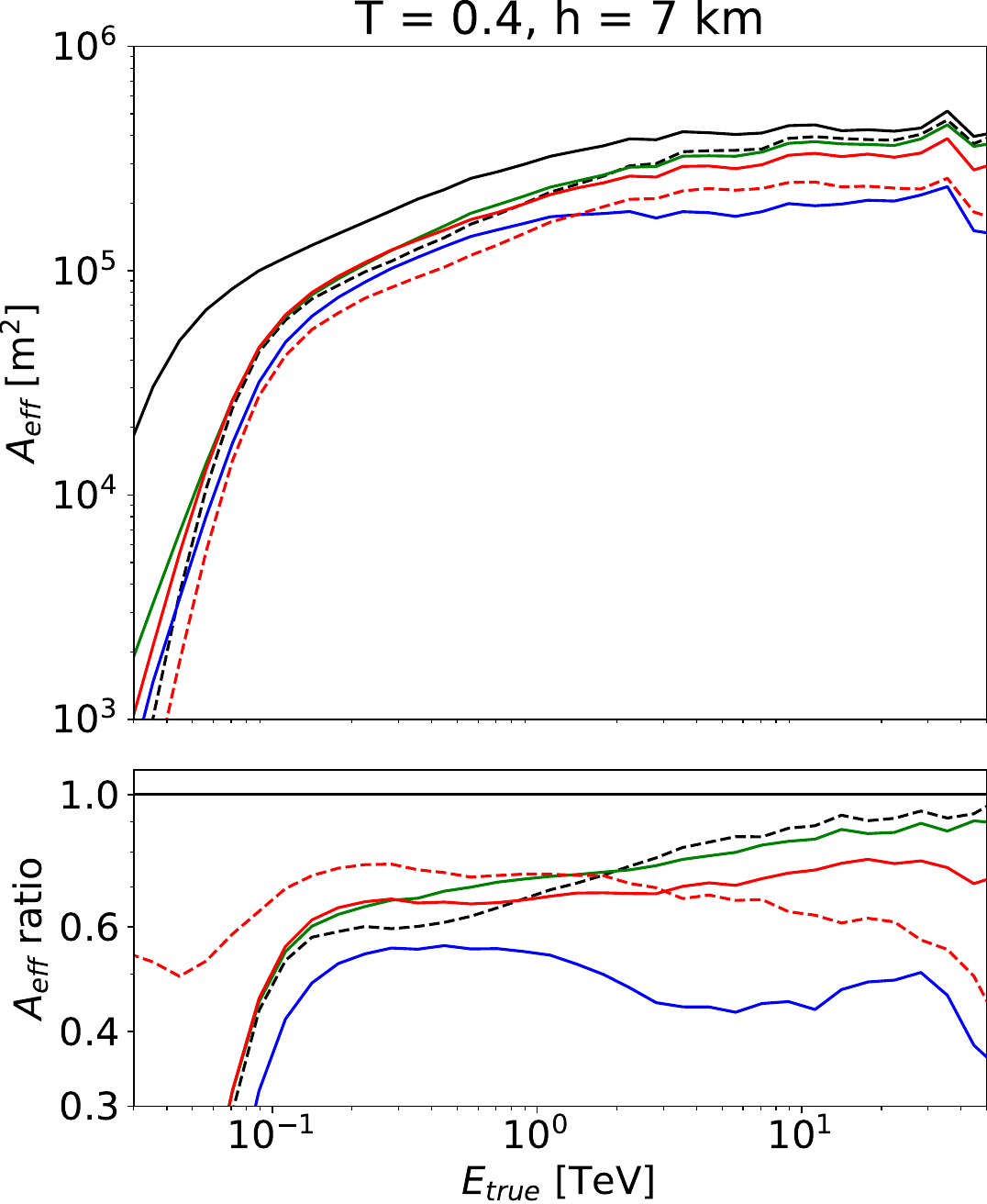}
    \includegraphics[width=0.33\textwidth]{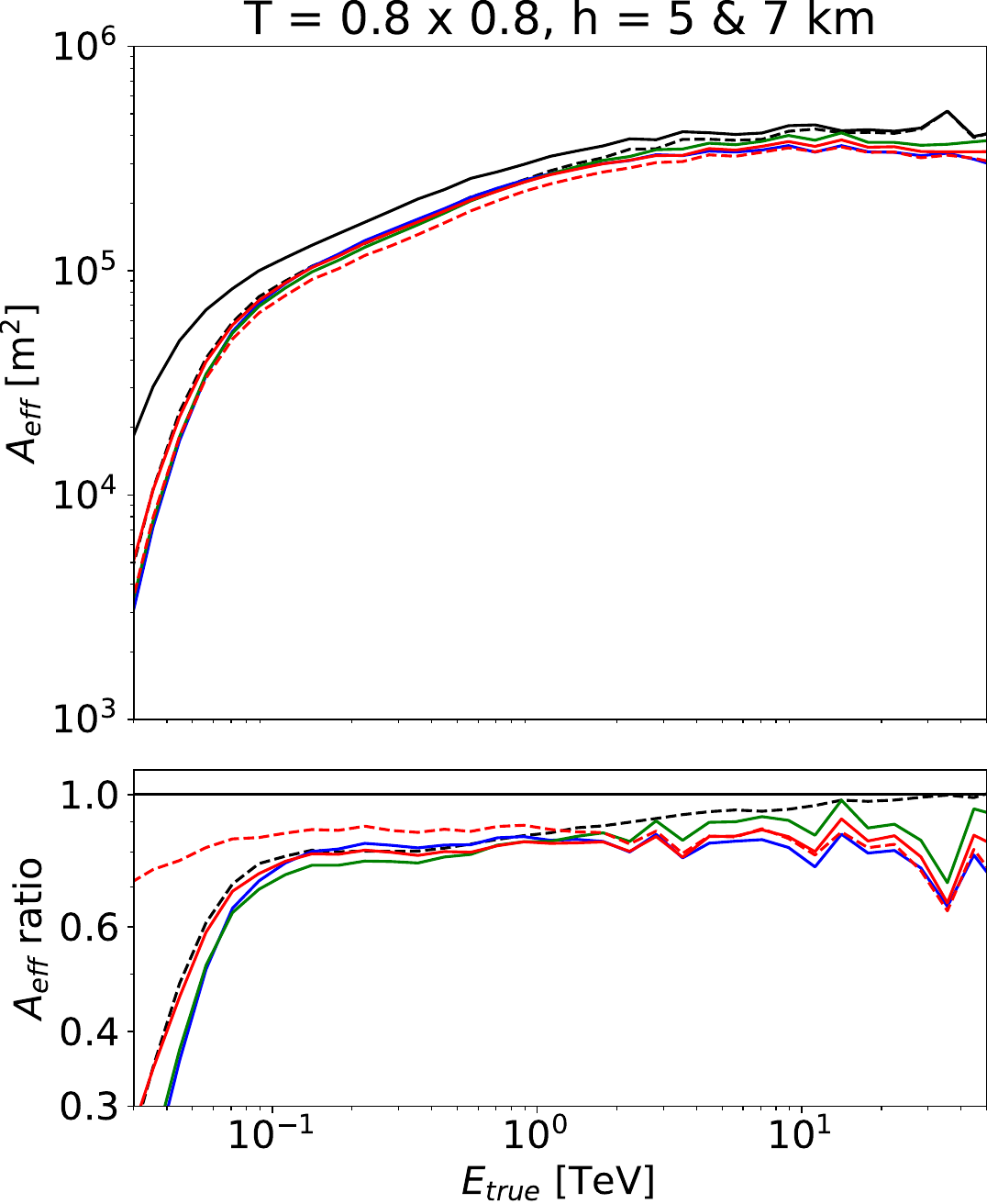}
    \caption{Collection area for gamma rays, expressed as a function of the true energy.
    Black lines: cloudless condition (T1), blue: cloudy data analysed with cloudless MC simulations, green: cloud analysed with dedicated MC simulations, red: image correction method analysed with general cloudless MC simulations.  
    Dashed lines show the results with an additional cleaning (AC).
    Different panels show different transmission, $T$, and base height, $h$, of the cloud.
    Efficiency cuts in \textit{gammaness} (90\%) and $\theta^2$ (70\%) are applied. 
    The second and fourth row sub-panels show selected ratios of the collection areas shown in the first and third row (see legend).
    }
    \label{fig:Aeff}
\end{figure*}

The results are shown in Fig.~\ref{fig:Aeff}. 
As expected (c.f. e.g. \citealp{2020APh...12002450S, 2023arXiv230202211P}), the presence of the cloud leads to the reduction of the collection area (see the blue curves in Fig.~\ref{fig:Aeff}). 
The effect is mostly pronounced at the lowest energies, where the collection area can drop by a factor of a few due to the increase of the energy threshold caused by stronger light extinction. 
At higher energies, the drop of the collection area is related mainly to the worse \textit{gammaness} evaluation and angular reconstruction due to inconsistency between the cloudless simulations used for the analysis of cloud-affected data. 
Depending on the cloud parameters, the effect (above the energy threshold) ranges from a few per cent for a $T=0.8$, $h=7$~km a.g.l. cloud up to $\sim$ 50\% for $T=0.8$, $h=7$~km a.g.l. 
Usage of dedicated MC simulations (see the green curves) including the cloud attenuation (following \cite{2023arXiv230202211P} approach) for the generation of RF models and calculation of the efficiency of cuts allows a partial recovery of the collection area, especially at high energies. 
This is related to the large images that are distorted by the cloud and therefore do not survive gamma/hadron separation cuts if based on training with cloudless simulations.
Interestingly, the method of image correction proposed here (shown in red curves) increases the achieved collection area compared to uncorrected data, to levels comparable to the one derived with MC simulations including a cloud. 
At medium and high energies the collection areas derived after correction are only $<20\%$ smaller than those obtained with cloudless simulations for most of the cases. As these values are smaller than typical systematic uncertainties of the current generation of Cherenkov telescopes, $\sim30\%$ (e.g. \citealp{2016APh....72...76A}, see also \citealp{2015ApJ...815L..22A}), it can be concluded that the proposed method can reproduce the spectra in the bulk of the IACT energy range.
For the most opaque clouds (at $T=0.4$) the underestimation is larger, but still $\sim30\%$ (lack of correction in this case results in underestimation of the flux by a factor of two). 

The second approach tested, which involves the application of an additional image cleaning both to the data and the simulations (dashed curves in Fig.~\ref{fig:Aeff}), provides even better agreements. 
Compared to the cloudless case (black dashed lines), at medium and high energies it results in only $\sim10\%$ underestimation of the collection area for most of the considered clouds (see the red dashed curves).
Similarly to the case of standard image cleaning, for the $T=0.4$ cloud, the underestimation is more pronounced, $\sim25\%$. 
Even more remarkably, such an analysis also recovers most of the collection area at the lowest ($\sim$ trigger threshold) energies resulting in only somewhat higher ($\sim20-30$\%) underestimations of the collection area. 
However, the additional cleaning approach also increases the energy threshold of the analysis (see the black line in Fig.~\ref{fig:Aeff}) by a factor of about $1/T$.

\subsection{Energy resolution and bias}
\label{sec:eres}

The energy reconstruction of IACTs is typically characterised by two quantities: the bias, which indicates the shift of energy estimation, and the resolution, which corresponds to the spread of the energy estimation at particular true energies.
We define the bias as the mean of the $(E_{est} / E_{true}) - 1$ distribution.
Since in one of the presented analyses we apply the energy estimation based on cloudless MC simulations to simulations including a cloud, the energy resolution requires additional attention.
In such a situation a significant bias can appear, and the energy estimation would be underestimated. 
Assuming that the underestimation is similar for all events of the same true energy, the whole distribution of the estimated to true energy ratio shifts to lower energies and shrinks by the factor of 1+bias.
Such an effect could be misinterpreted as an improvement of the energy resolution. 
To counteract this effect we therefore evaluate the energy resolution corrected by the 1+bias factor. 

\begin{figure*}[tp!]
    %\centering
    \includegraphics[width=0.33\textwidth]{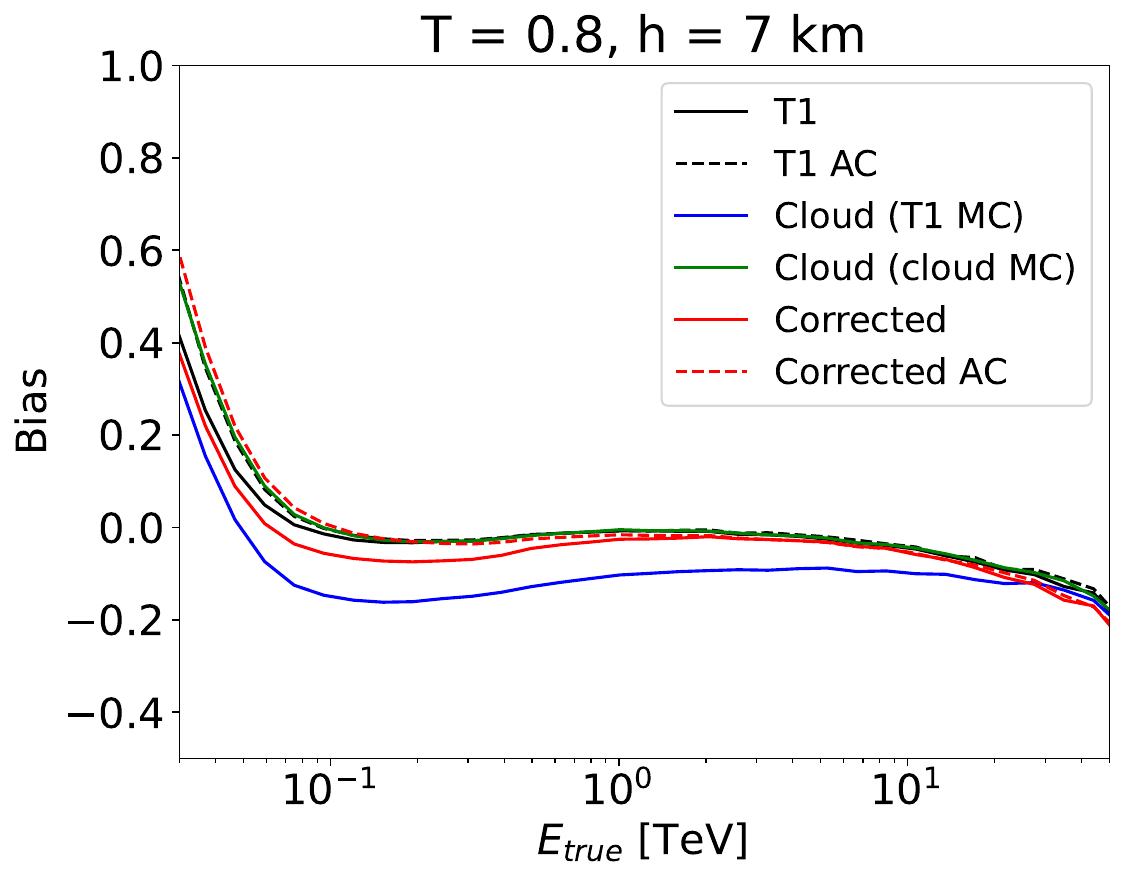}
    \includegraphics[width=0.33\textwidth]{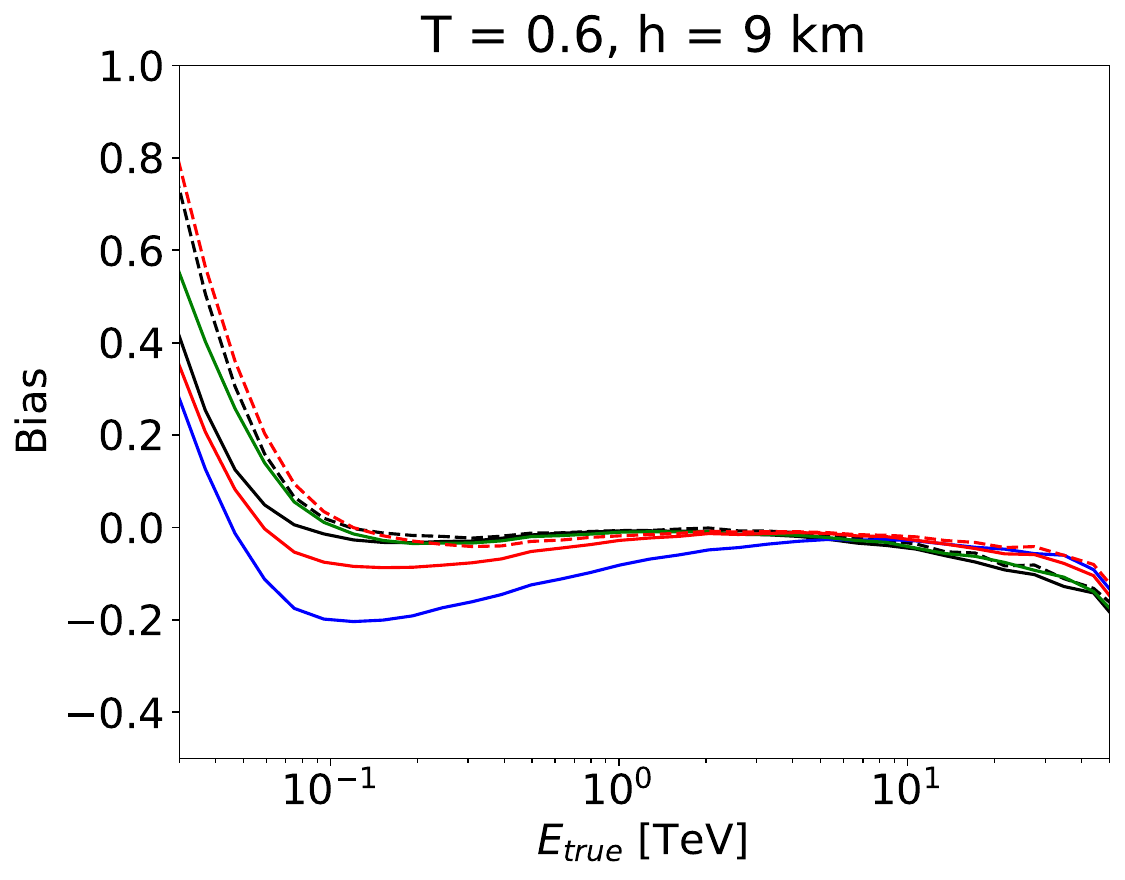}
    \includegraphics[width=0.33\textwidth]{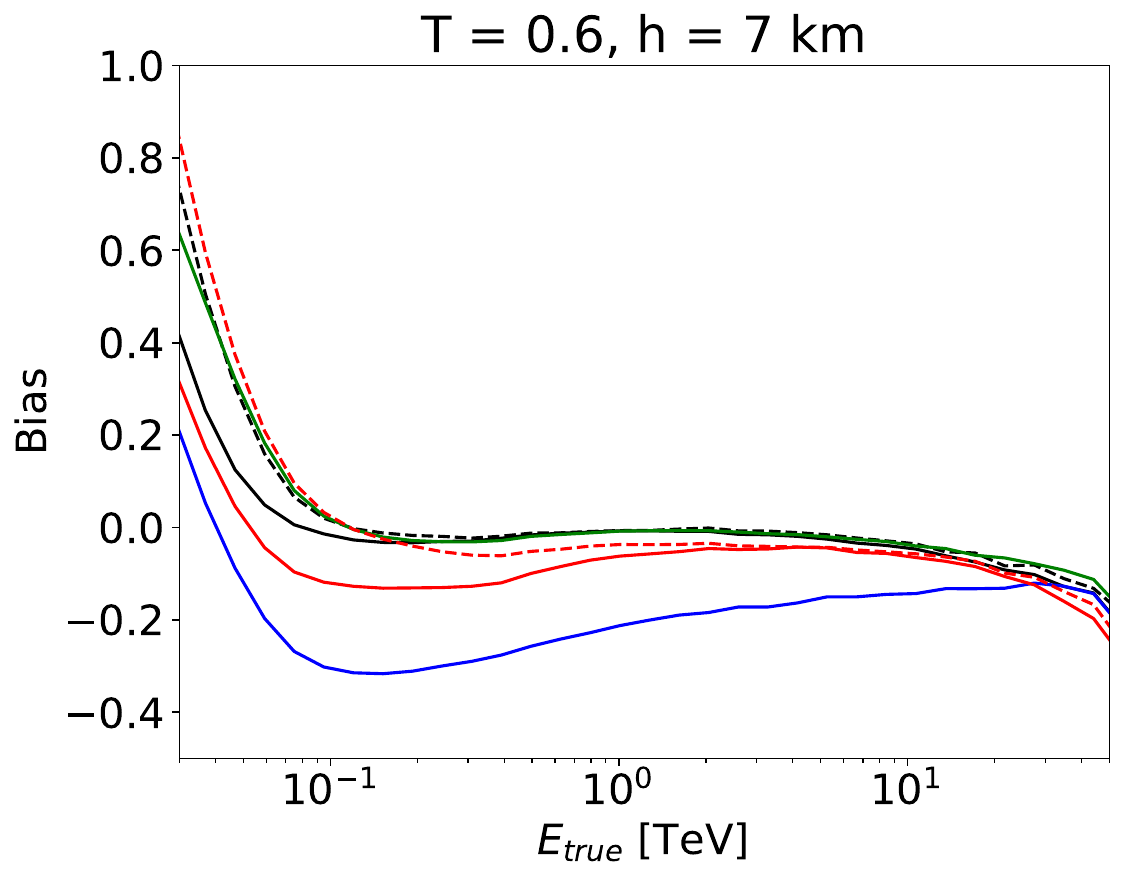}
    \includegraphics[width=0.33\textwidth]{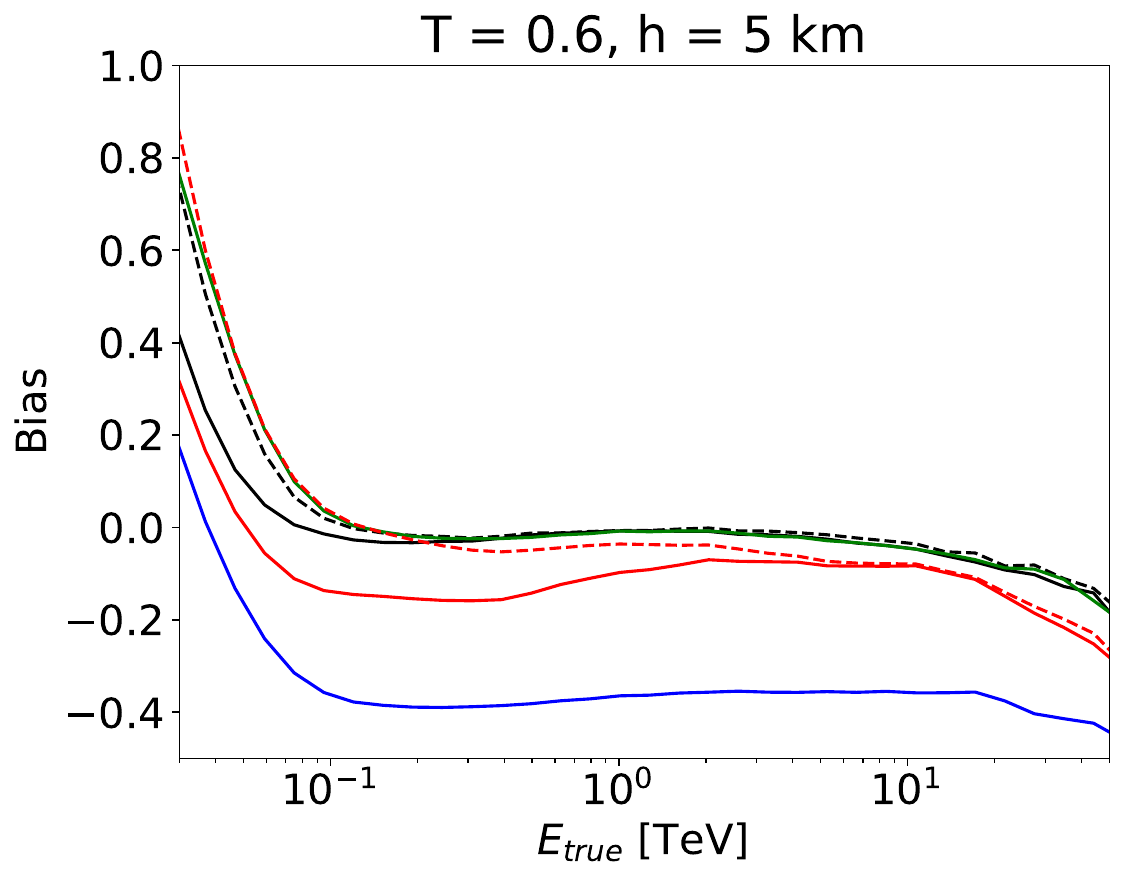}
    \includegraphics[width=0.33\textwidth]{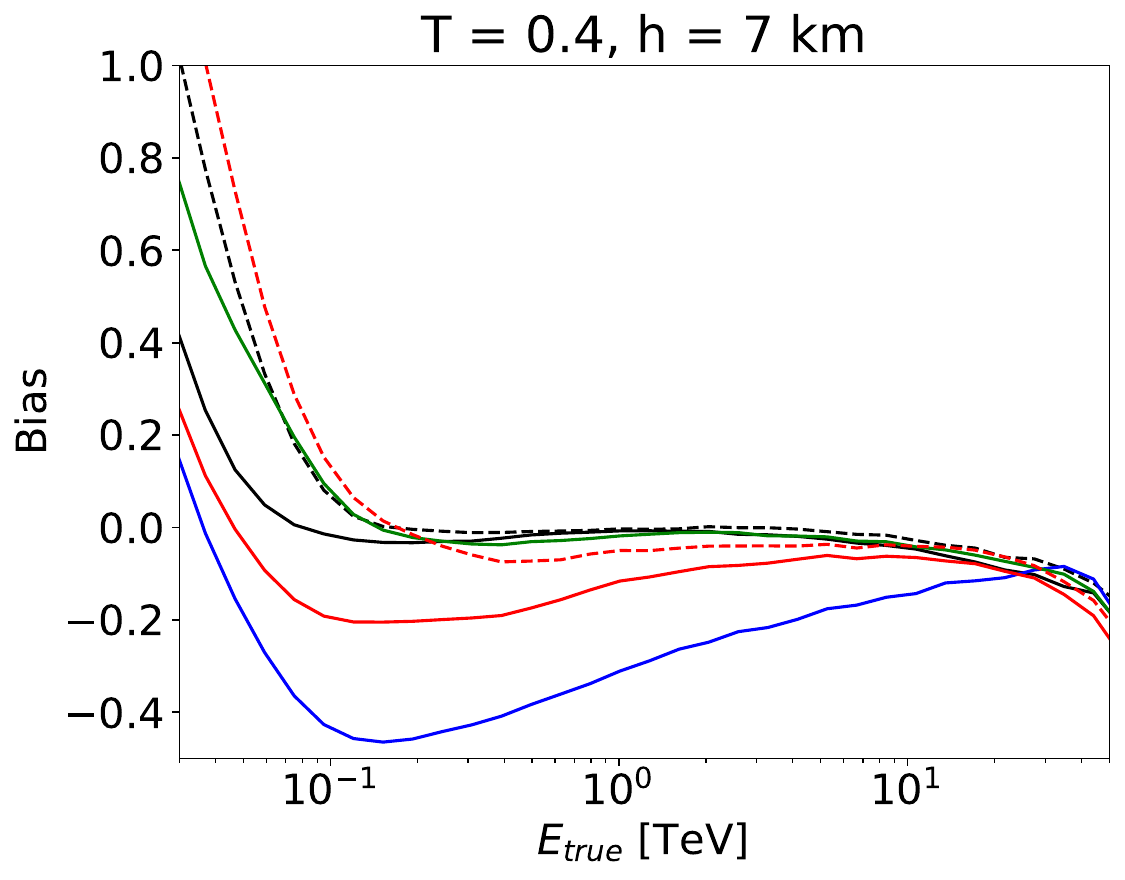}
    \includegraphics[width=0.33\textwidth]{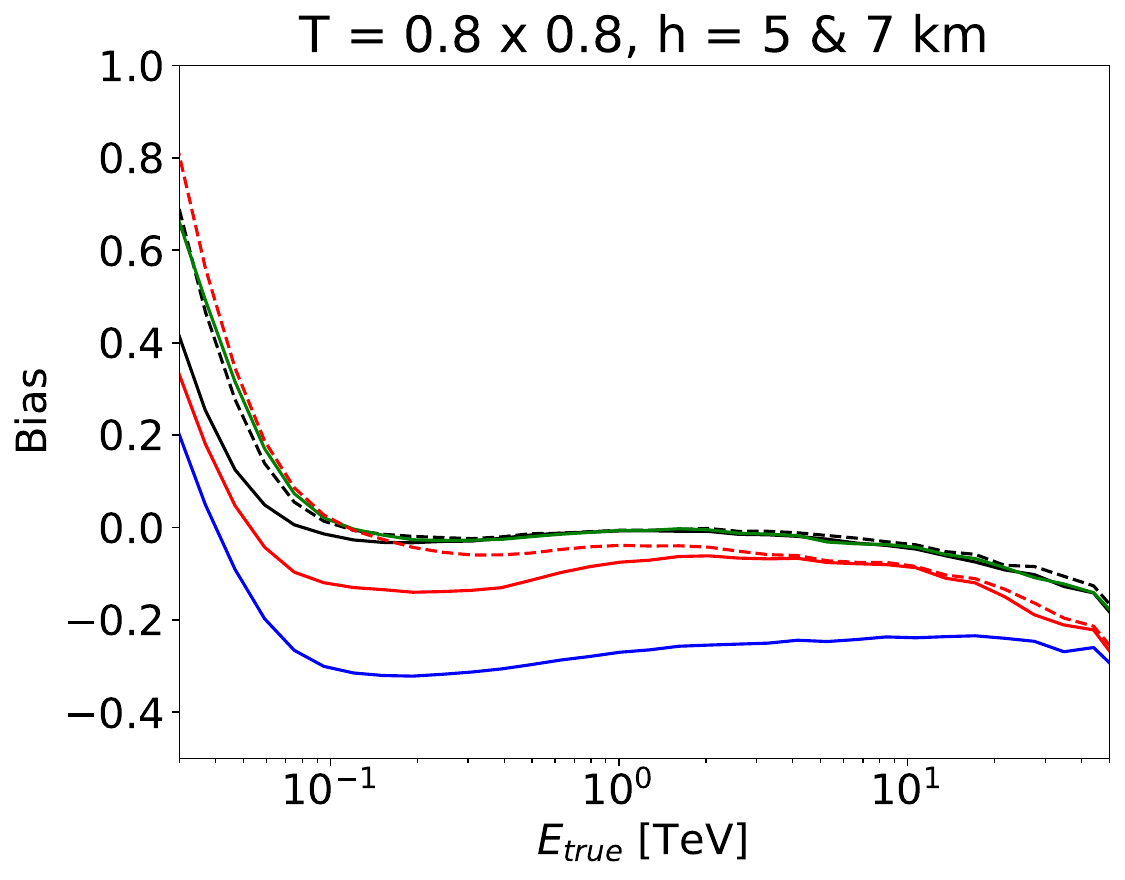}
    \caption{Comparison of energy bias for different simulations and analyses.
    Black lines: cloudless condition (T1), blue: cloud analysed with cloudless MC simulations, green: cloud analysed with dedicated MC simulations, red: image correction method analysed with general cloudless MC simulations.  
    Dashed lines show the results with an additional cleaning (AC).
    Different panels show different transmission, $T$, and base height, $h$, of the cloud.}
    \label{fig:ebias}
\end{figure*}

\begin{figure*}[tp!]
    %\centering
    \includegraphics[width=0.33\textwidth]{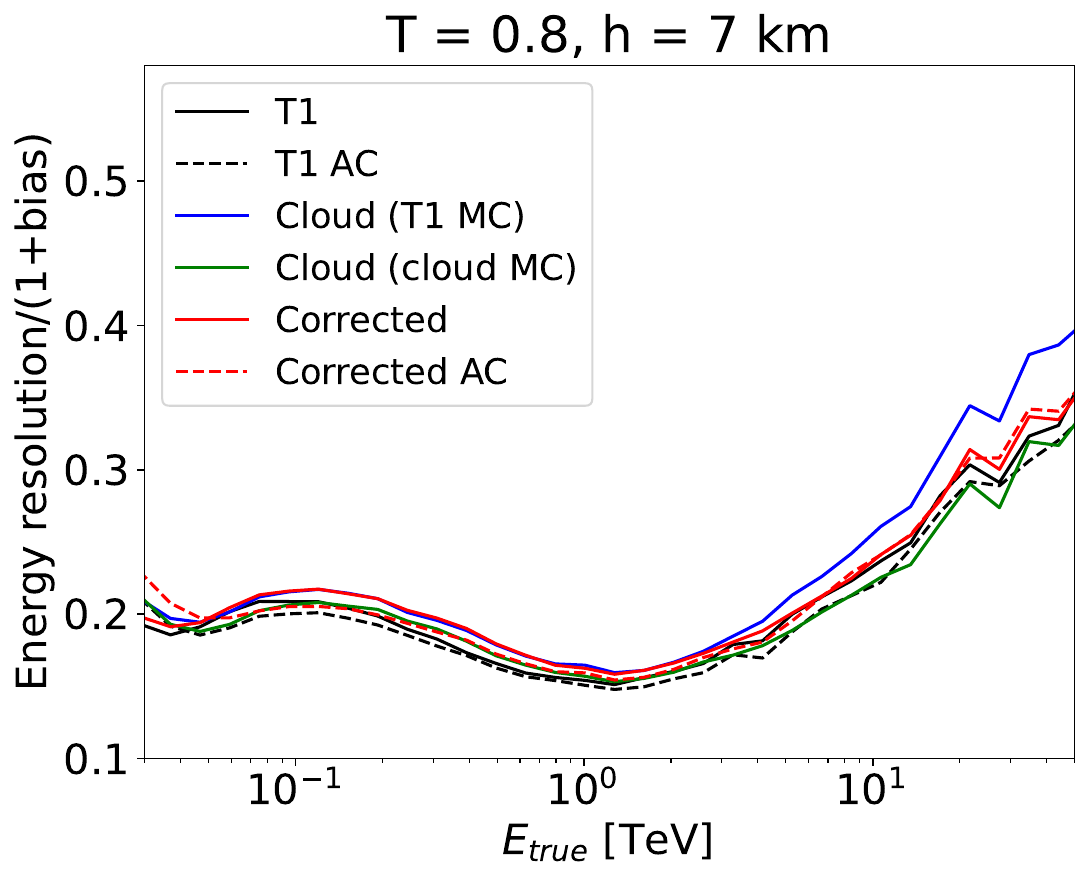}
    \includegraphics[width=0.33\textwidth]{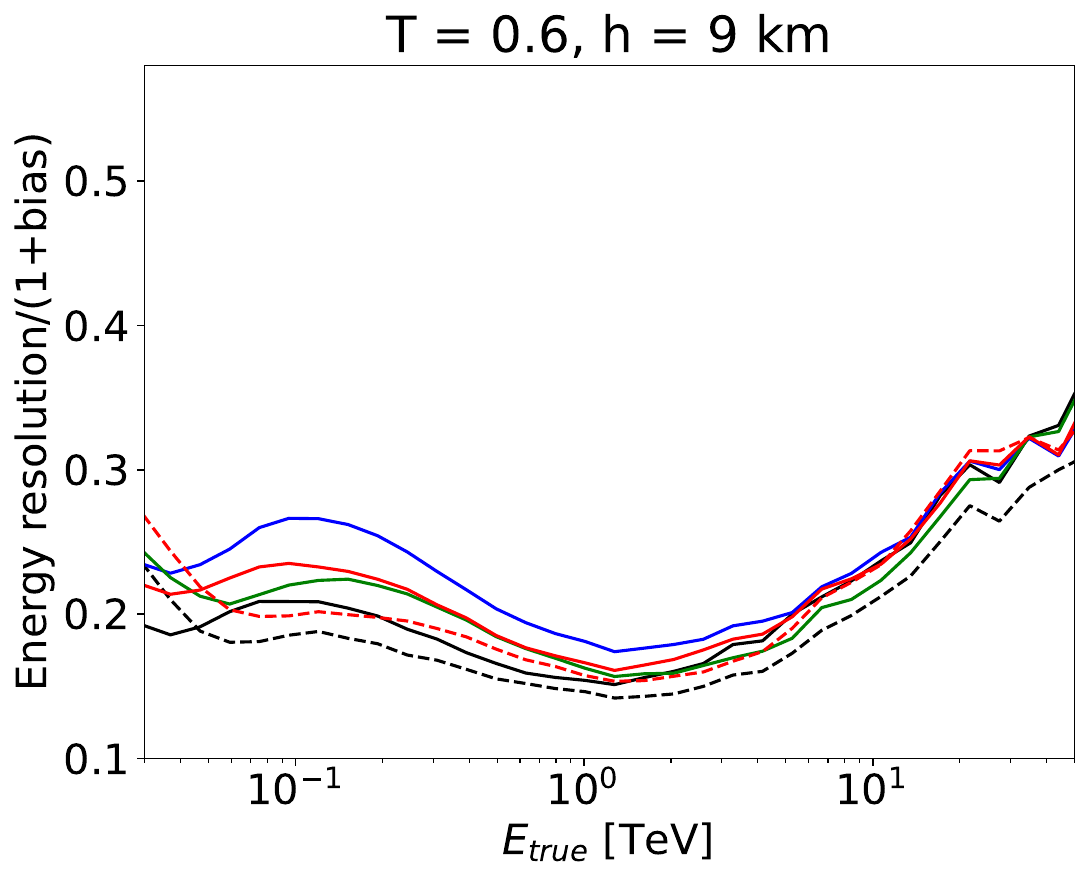}
    \includegraphics[width=0.33\textwidth]{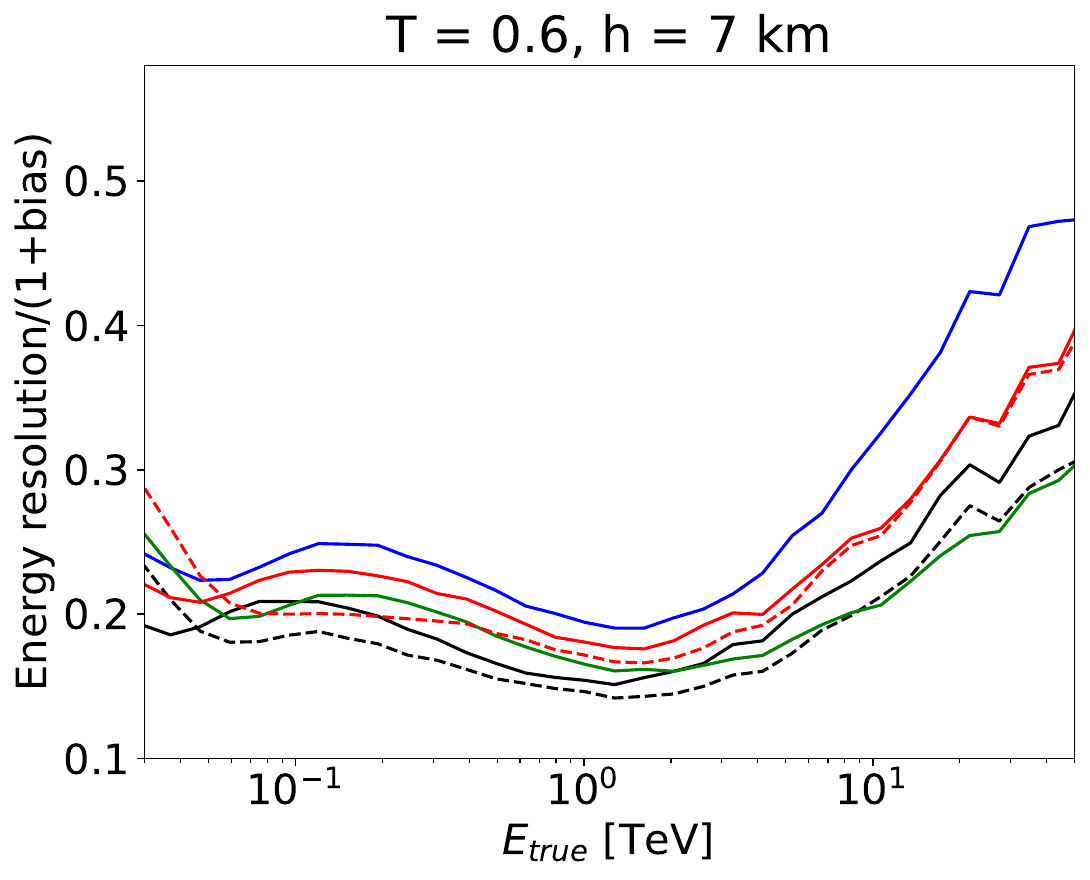}\\
    \includegraphics[width=0.33\textwidth]{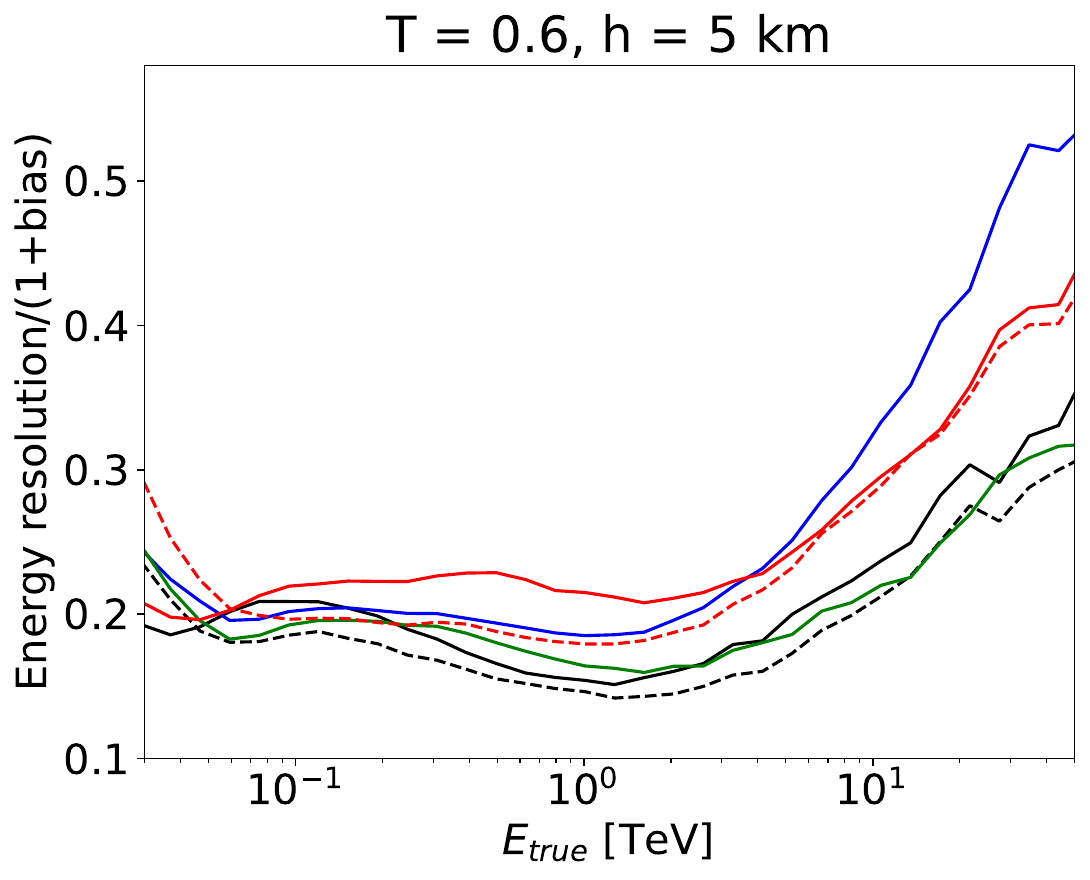}
    \includegraphics[width=0.33\textwidth]{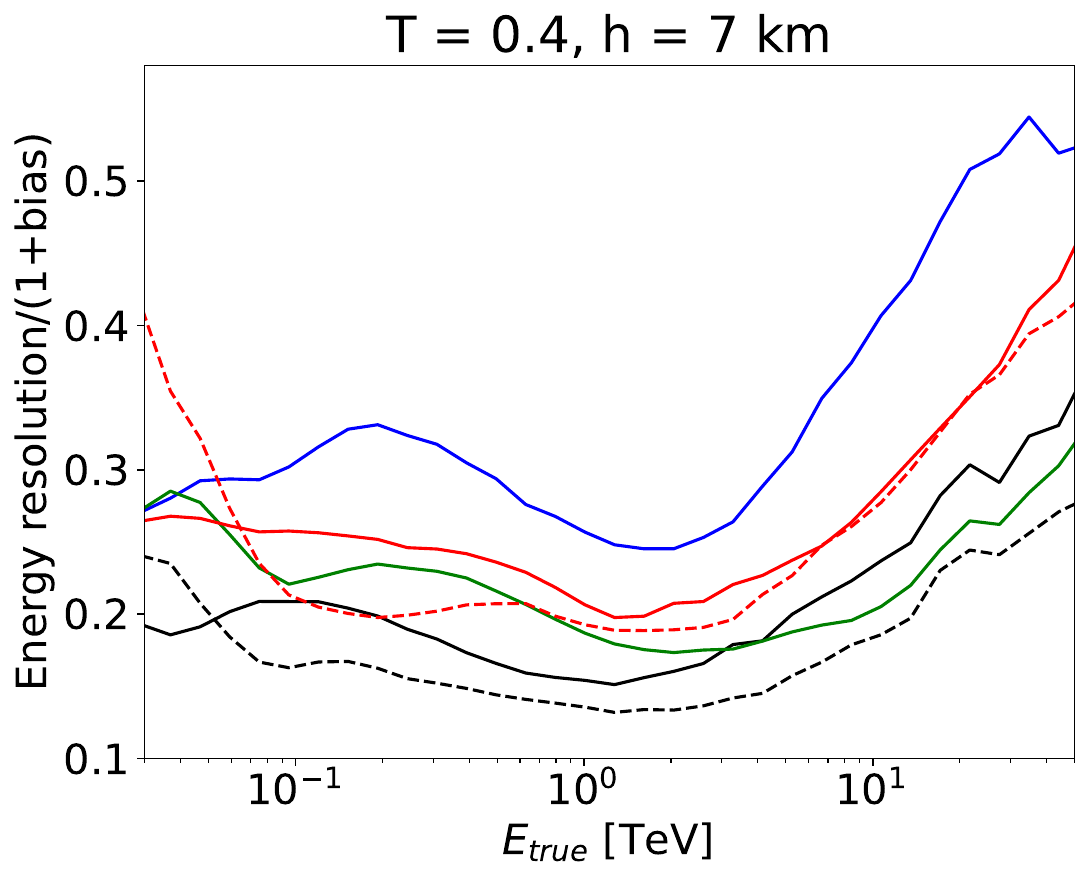}
    \includegraphics[width=0.33\textwidth]{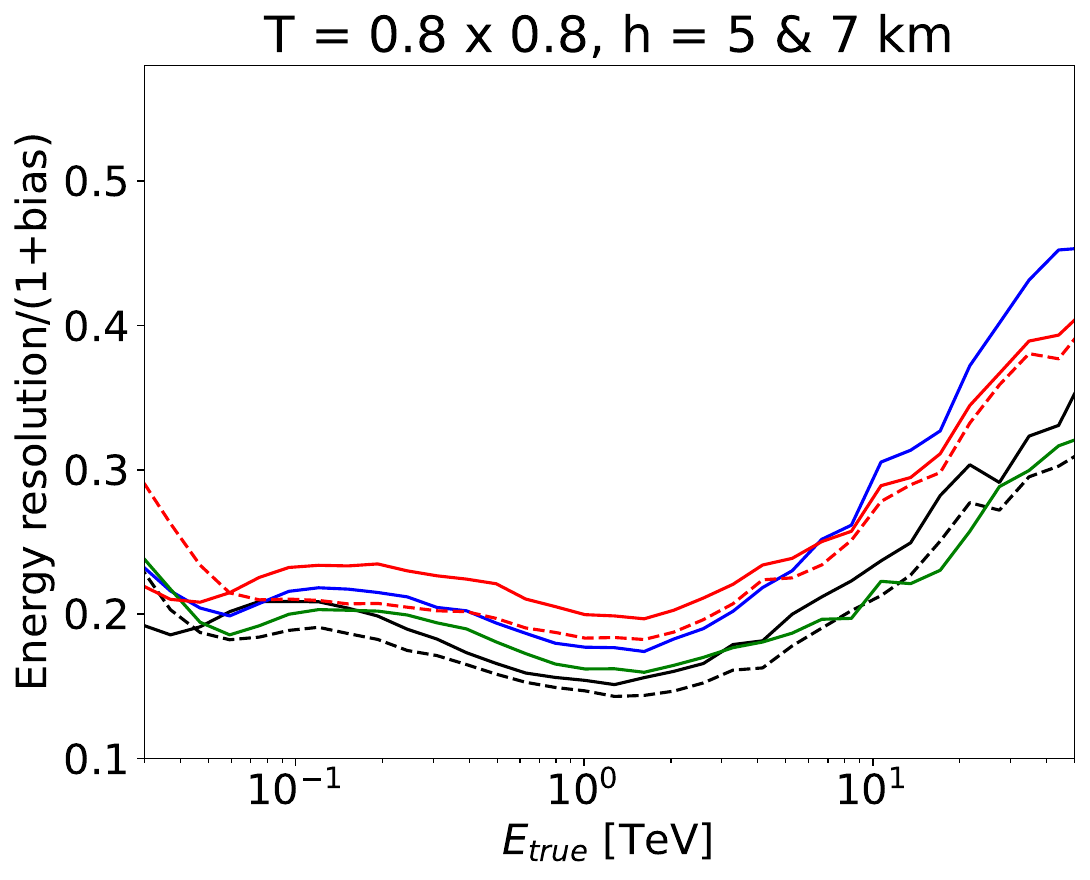}
    
    \caption{Comparison of energy  resolution (corrected for 1+bias factor) for different simulations and analyses (lines and panels like in Fig.~\ref{fig:ebias})}
    \label{fig:eres}
\end{figure*}

The results are shown in Fig.~\ref{fig:ebias} and Fig.~\ref{fig:eres}.
%\komm{here the biases are important - they show if we are correcting the effect properly}
%
We reproduce the effect of the energy bias introduced by the cloud seen in earlier studies (see e.g. \citealp{Nolan10f1,2020APh...12002450S}). 
The negative bias is the largest at low ($\sim100$~GeV) energies and decreases at higher energies as a bigger fraction of the shower is developing below the cloud (see blue curves in Fig.~\ref{fig:ebias}). 
In the case of low clouds ($h=5$~km a.g.l.), even the showers initiated by $\sim 10$~TeV energies develop above the cloud resulting in a nearly constant bias. 
As expected, the bias is removed (except for the threshold effect at the lowest energies) if consistent cloud-affected MC simulations are used to train the energy estimation (see the green curves). 
The proposed method of image correction (see the red curves) can remove most of the energy bias without the need for generating special MC simulations.
Only for the $T=0.6$ at $h=5$~km a.g.l. and $T=0.4$ at $h=7$~km a.g.l. clouds a considerable residual bias of up to 15\% -- 20\% remains. 
With the usage of the additional image cleaning, at medium and high energies, the bias is nearly completely removed (below a few per cent) even for the most opaque cloud. 
At the lowest energies, affected by the threshold effects, the bias is larger. 
The increase of the bias at the lowest energies with the usage of additional image cleaning is caused by the increase in the energy threshold. 

The (corrected for bias) energy resolution is also degraded by the presence of the cloud.
As expected, the effect is largest if no treatment for the presence of the cloud is applied either by using specialised MC simulations or by data correction (see the blue curves in Fig.~\ref{fig:eres}). 
The energy resolution in this case can be as bad as $40-50\%$ at $\sim 10$~TeV energies for more opaque or lower clouds.
The effect is however reduced if either dedicated MCs are applied, or the images are corrected using the proposed method. 
For the cases in the top row of Fig.~\ref{fig:eres}, the proposed image correction method (red curves) results in a very similar energy resolution as obtained using dedicated cloud MC simulations (green curves). 
The complicated behaviour of the curves at energies of tens of GeV is caused by the strong energy bias.

The use of additional cleaning (dashed curves) improves the energy resolution, also in the case of lack of a cloud. 
This is expected since such cleaning will remove small images, mostly produced by high-impact events, which have worse energy resolution.
The same effect is likely also responsible for the energy resolution at $\sim10$~TeV becoming slightly better due to the presence of a cloud. 
On the other hand, the additional cleaning does not reduce the small mismatch in the energy resolution between cloudless simulations and corrected cloud images. 

\subsection{Angular resolution}
\label{sec:angres}

\begin{figure*}[tp!]
    %\centering
    \includegraphics[width=0.33\textwidth]{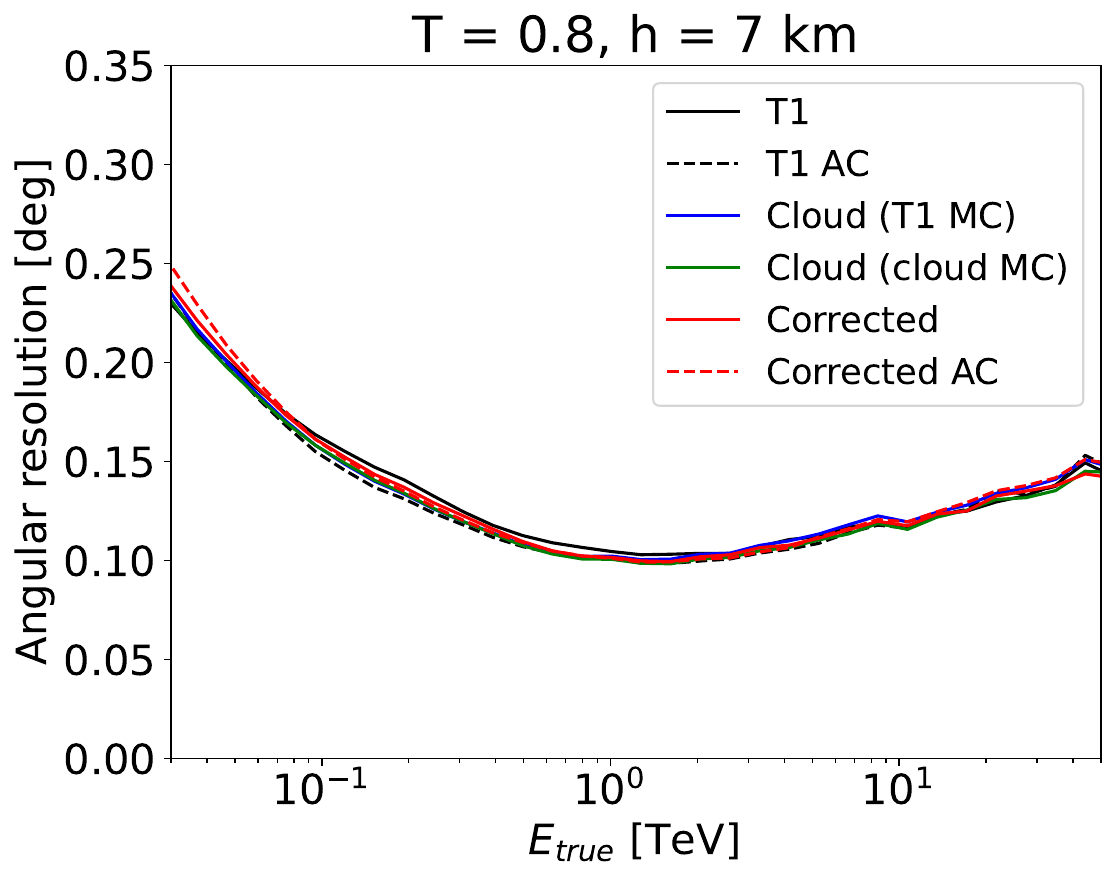}
    \includegraphics[width=0.33\textwidth]{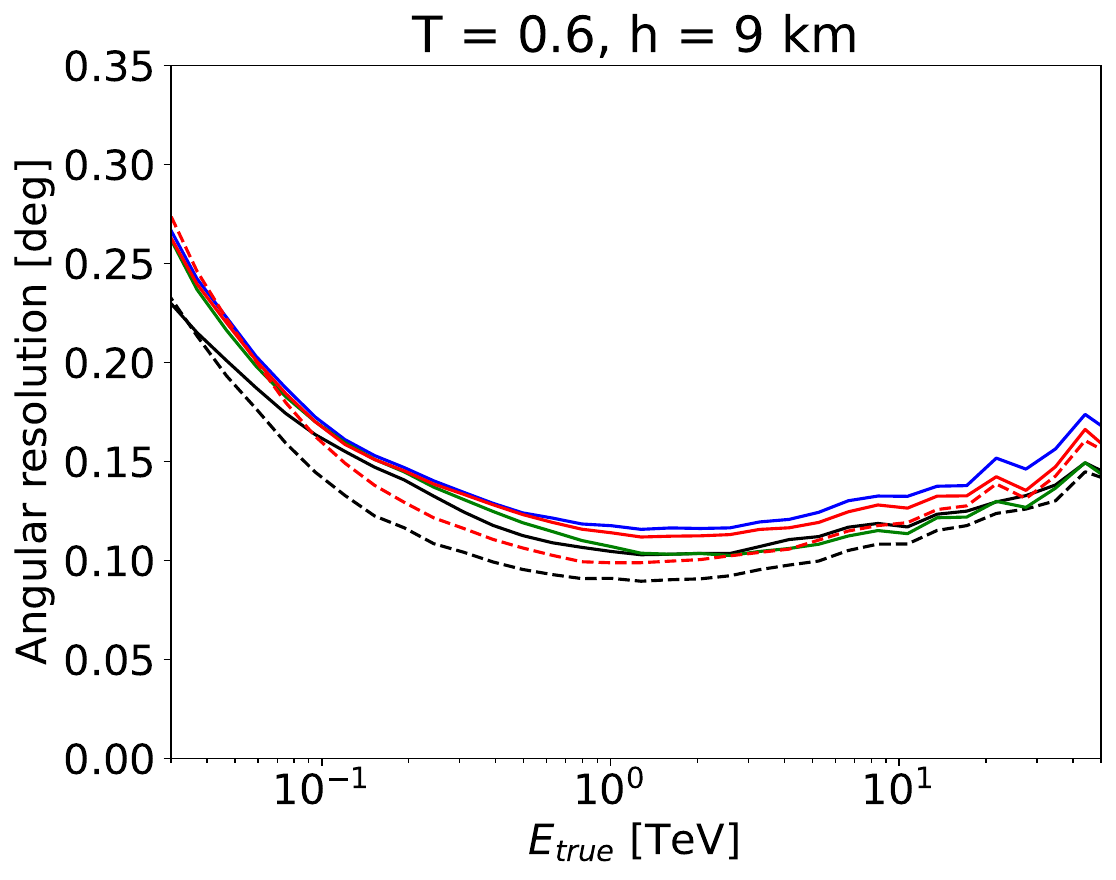}
    \includegraphics[width=0.33\textwidth]{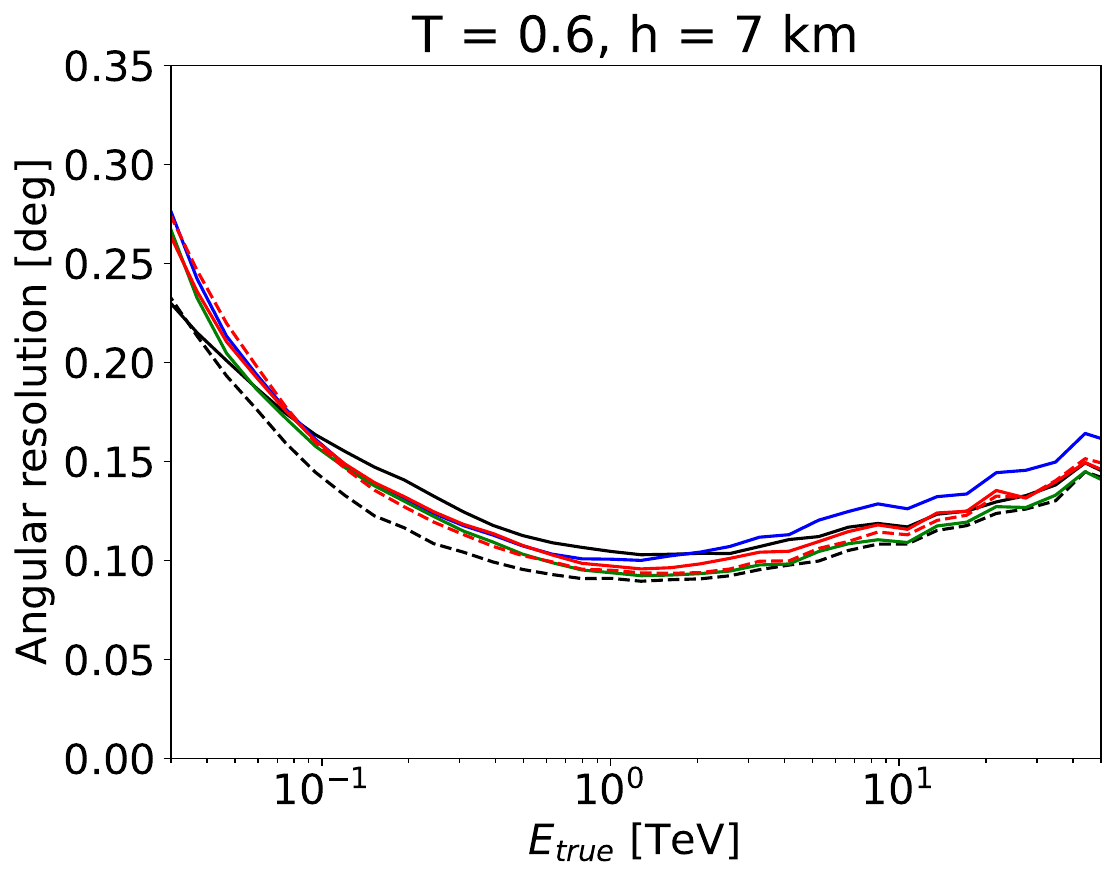} \\
    \includegraphics[width=0.33\textwidth]{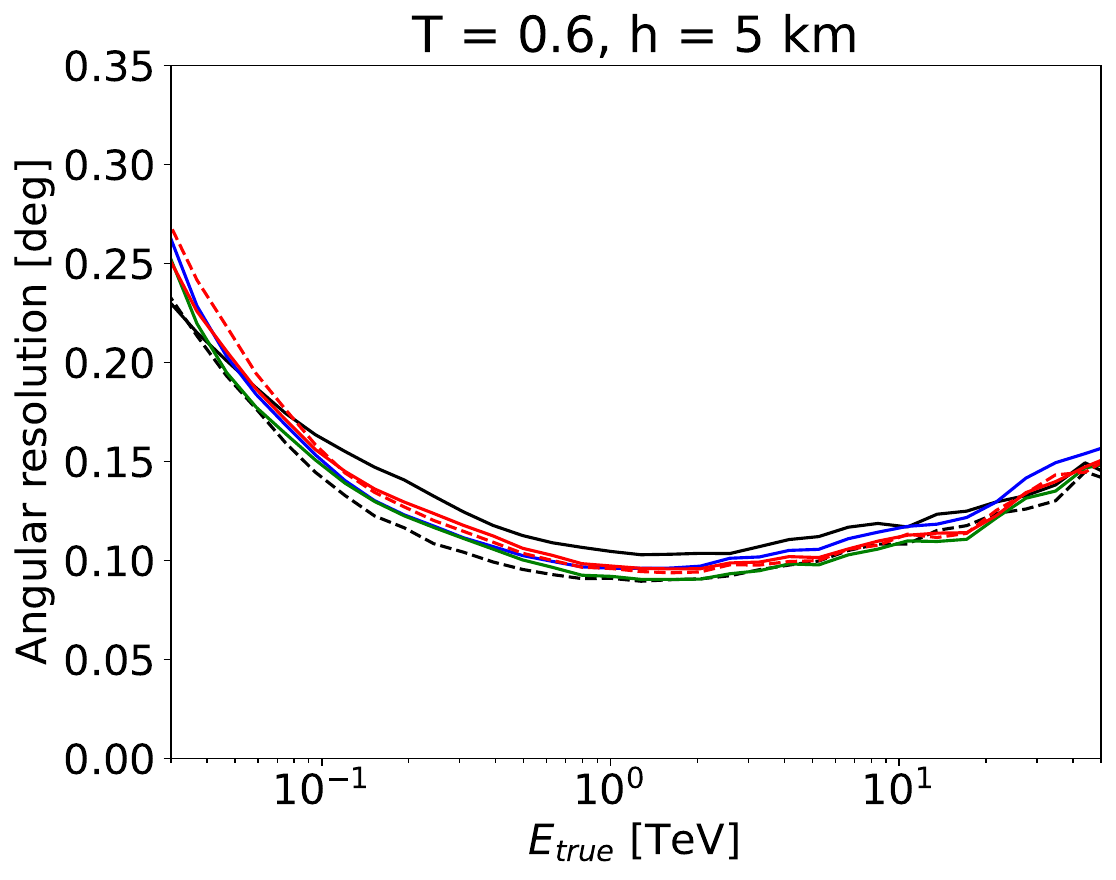}
    \includegraphics[width=0.33\textwidth]{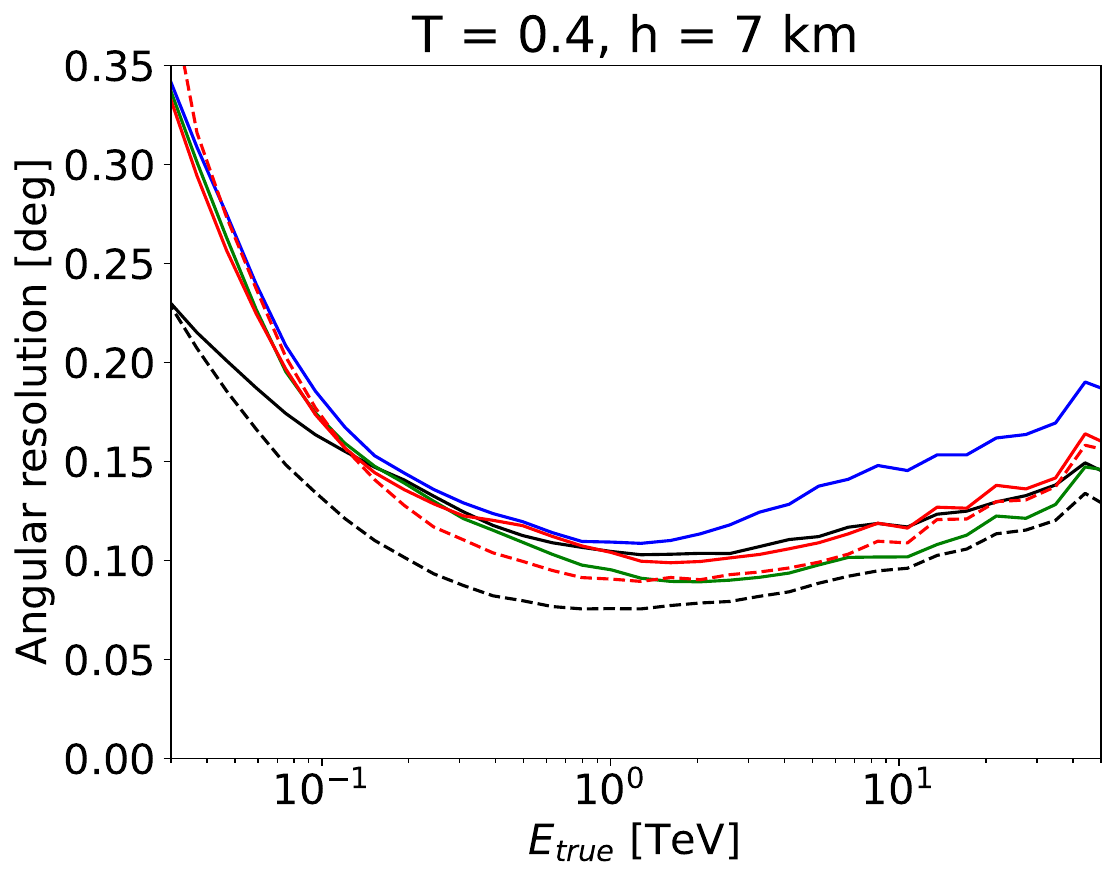}
    \includegraphics[width=0.33\textwidth]{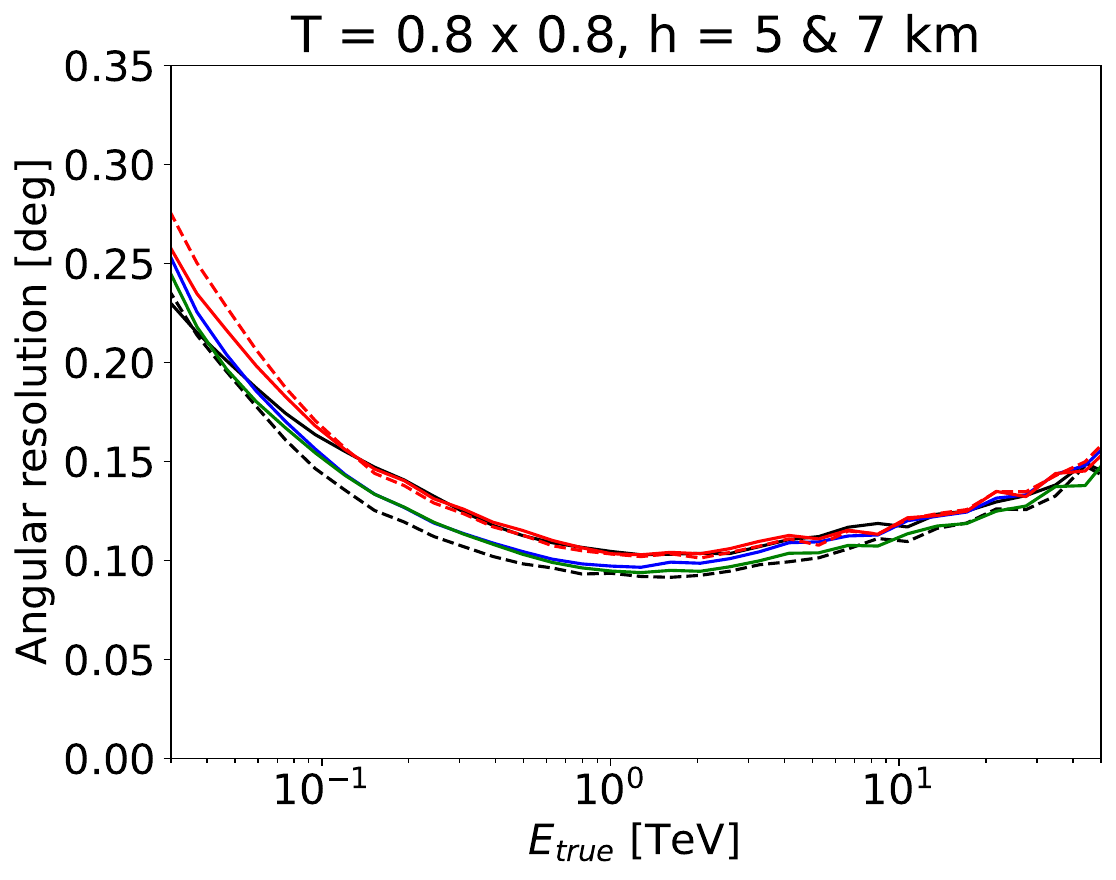}
    \caption{Angular resolution for cloudless simulations and clouds with different transmission, $T$, and base height, $h$. 
    Black lines: cloudless condition (T1), blue: cloud analysed with cloudless MC simulations, green: cloud analysed with dedicated MC simulations, red: image correction method analysed with general cloudless MC simulations.  
    Dashed lines show the results with an additional cleaning (AC).
}
    \label{fig:angres}
\end{figure*}

We define the angular resolution as the containment radius of 68\% of gamma rays in a particular energy bin.
To avoid issues with the energy bias the data are binned following their true energy. 
We summarise the results in Fig.~\ref{fig:angres}.
The angular resolution is the performance parameter least affected by the presence of clouds. 
At the lowest energies the light extinction results in a minor worsening of the angular resolution. 
This is also in line with the results of \citet{2020APh...12002450S}, where the angular resolution was modified by the cloud mostly by the threshold effect. 
Similarly to the case of energy resolution, the presence of a cloud in some cases can slightly improve the angular resolution at the highest energies, by removing weaker, harder to reconstruct images, leaving only closer events.
This makes a complicated dependence of the cloud parameters, as clouds of the same transmission but different base heights (compare the case of $T=0.6$, $h=7$~km a.g.l. with $h=9$~km a.g.l.) could produce a net improvement or net worsening of the angular resolution.
Nevertheless, for most of the cloud parameters and in the middle energy range the effects are $\lesssim10\%$. 
This is expected, as the angular reconstruction is mostly based on the direction of the main axis, which due to symmetry is not affected by the cloud. 
The bias in the image and stereoscopic parameters caused by the cloud is not large enough to strongly affect the resolution. 
In general, as expected, the use of cloudless MC simulations for the reconstruction of cloud-affected images worsens the angular resolution, however, the effect is at most mediocre.
The performance with the image correction method is in between the cases of lack of correction and usage of dedicated MC simulations.

Similar to the case of energy resolution, the use of additional cleaning can slightly improve the angular resolution in the cloudless case (by removing dim images). 
It does not improve however the relative performance of the correction method. 

\subsection{Sensitivity}
\label{sec:sens}

To evaluate the performance of the proposed method in the detection of gamma-ray sources, we derive the sensitivity for different types of clouds and corrections. 
We follow a commonly used definition of differential sensitivity, namely the flux that fulfils the following conditions: (1) results in $5\sigma$ significance according to Eq.~17 of \citet{1983ApJ...272..317L}, (2) provides at least 10 excess events and (3) exceeds 5\% of the residual background.
The sensitivity is derived in 5 bins per decade of energy for 50~hrs of observations. 
Contrary to the case of angular and energy resolution there is no straightforward way of computing the sensitivity in true energy bins\footnote{The reason for this is that for gamma rays with a given true energy the background is composed of protons, electrons and other particles of different distributions of their true energies.}. 
Therefore, the sensitivity is typically derived in bins of estimated energy.
This however causes a problem in the case of a strong energy bias (that will be the case if we compare to the analysis not applying the correction for the cloud presence). 
Thus, we first apply a simple correction for the bias, by computing the expected value of $b(E_{est})=(E_{est}-E_{true})/E_{true}$ in each estimated energy bin. 
We can then compute the bias-corrected estimated energy: $E'_{est} = E_{est}/(1+b(E_{est}))$. 

Next, to efficiently use the available background statistics, we use a k-fold cross-validation method. 
Namely, we divide the whole statistics into 4 sub-samples. 
In each sub-sample and each bin of $E'_{est}$ we derive the \textit{gammaness} and $\theta^2$ cuts maximising the sensitivity by optimising the cuts on the remaining three sub-samples. 
The final, unbiased sensitivity is obtained by stacking the gamma-ray and background rates from all sub-samples. 

\begin{figure*}[tp!]
    \centering
    \includegraphics[width=0.49\textwidth]{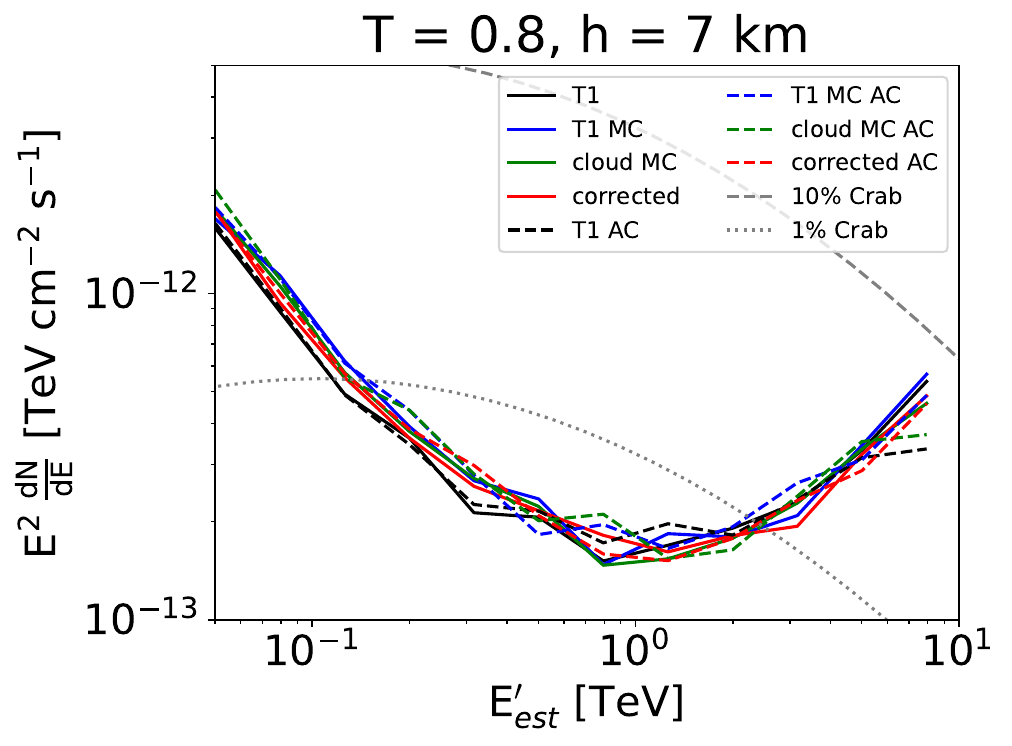}
    \includegraphics[width=0.49\textwidth]{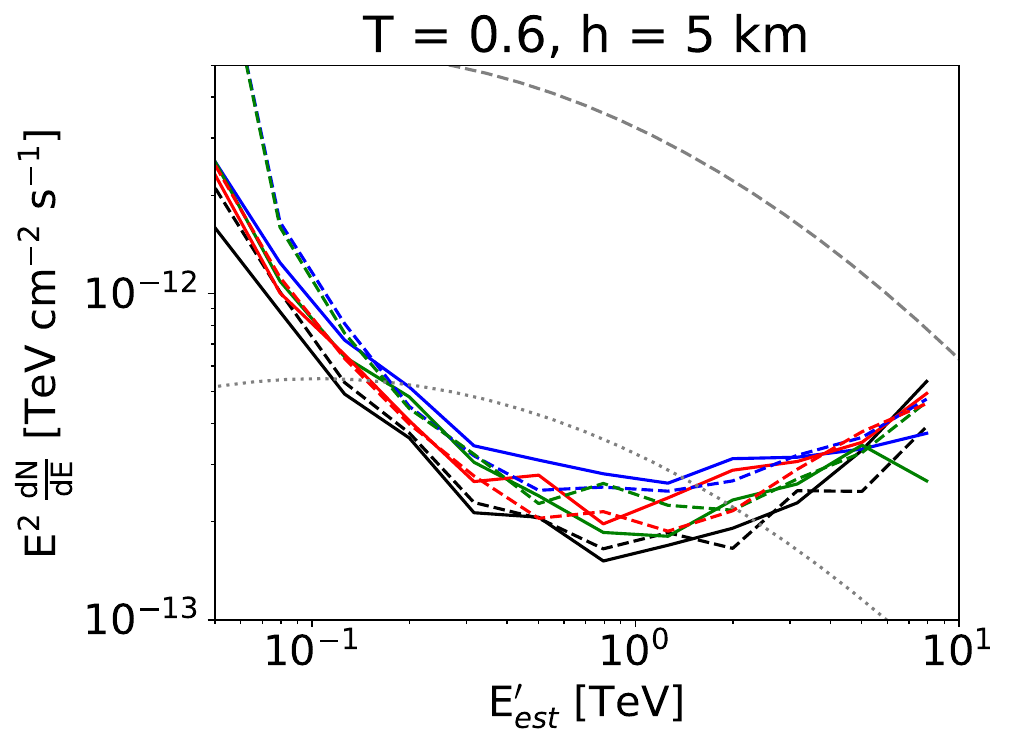}
    \includegraphics[width=0.49\textwidth]{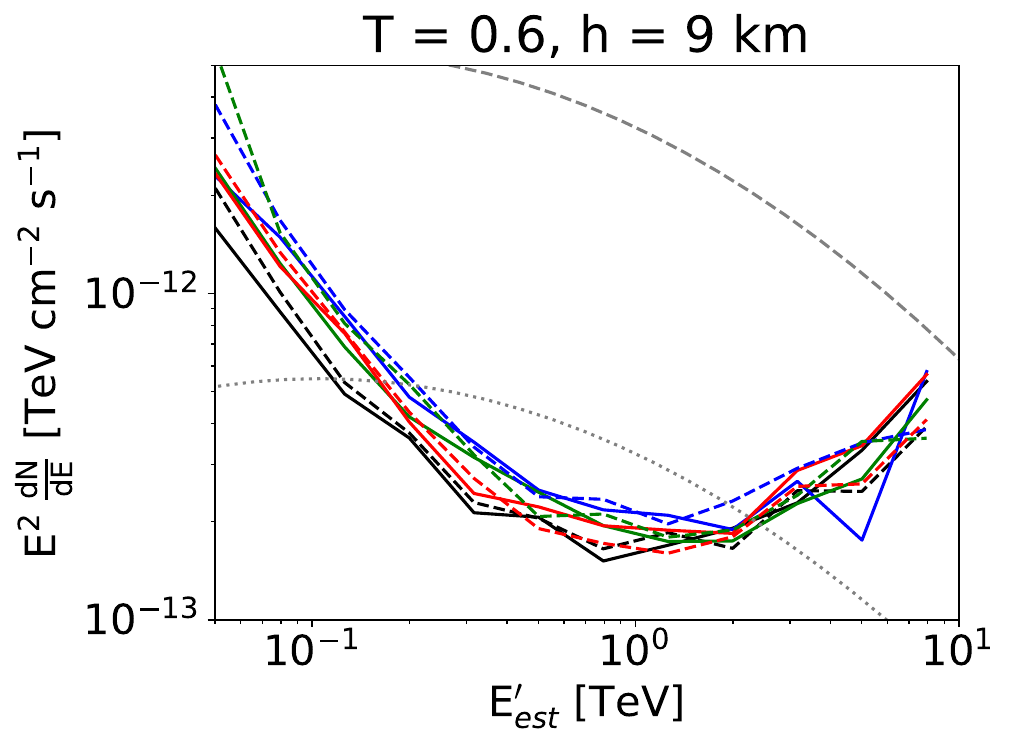}
    \includegraphics[width=0.49\textwidth]{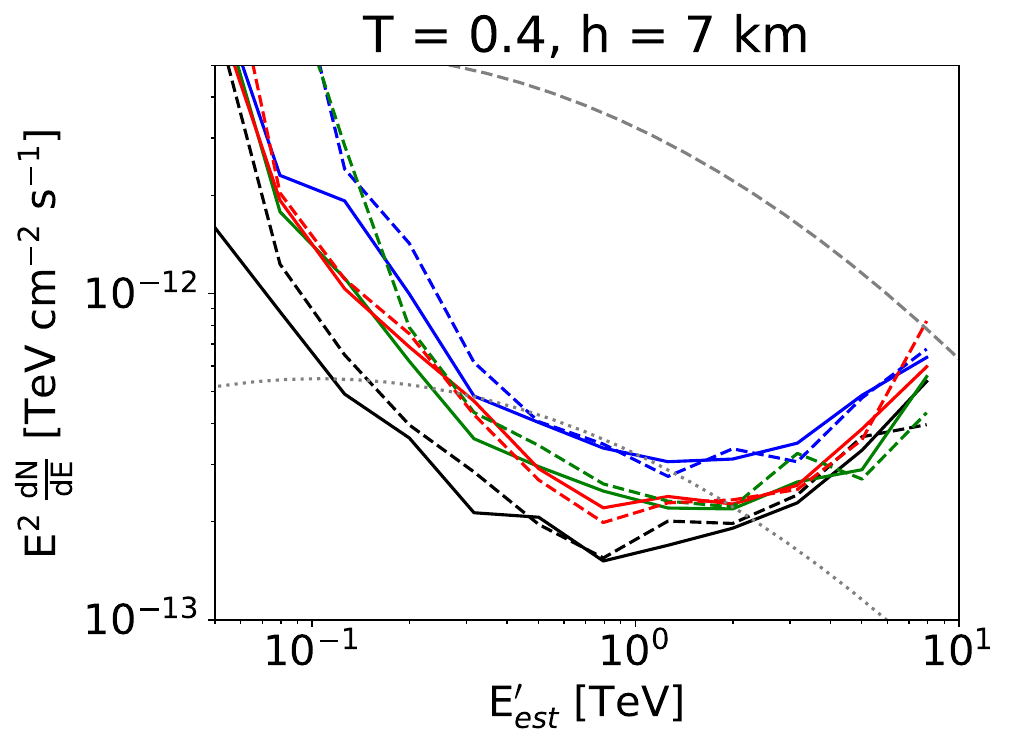}
    \includegraphics[width=0.49\textwidth]{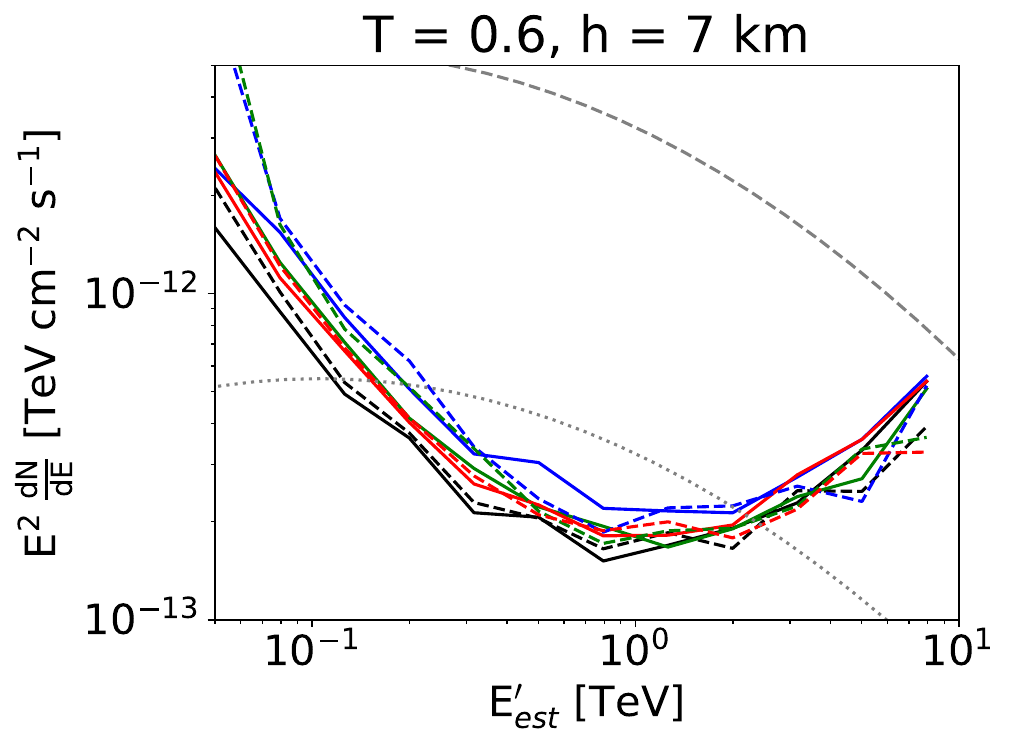}
    \includegraphics[width=0.49\textwidth]{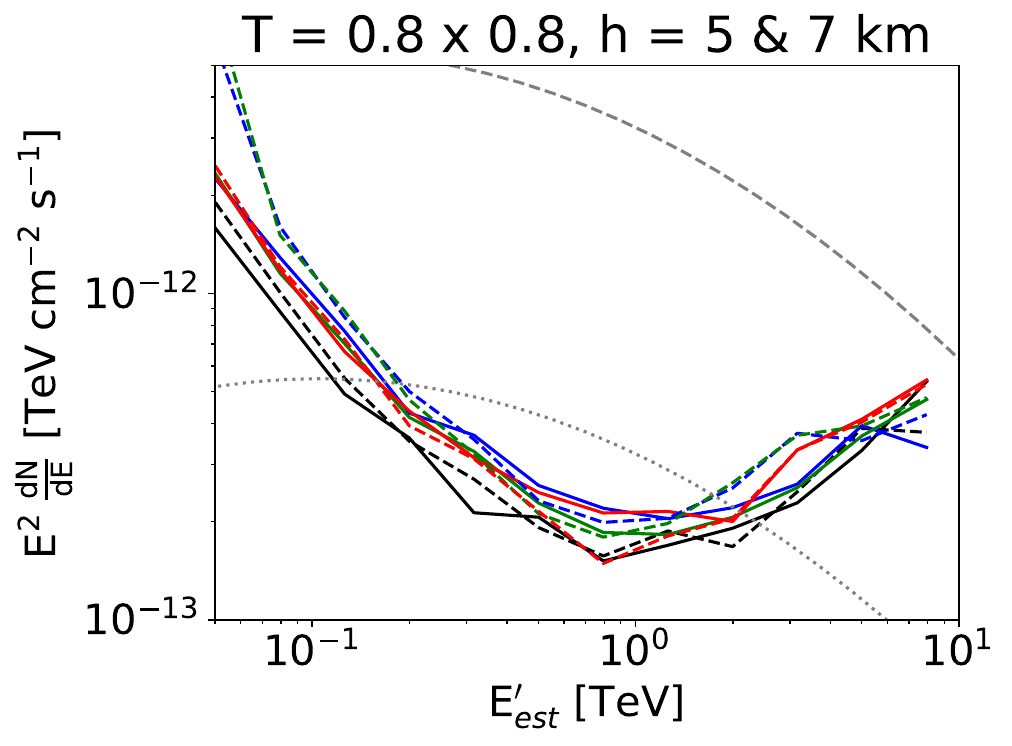}
    
    \caption{Comparison of sensitivity. 
    Black lines: cloudless condition (T1), blue: cloud (of the type listed above the figure) analysed with cloudless MC simulations, green: cloud analysed with dedicated MC simulations, red: image correction method analysed with general cloudless MC simulations.  
    Dashed lines show the results with an additional cleaning (AC).    
    }
    \label{fig:sens}
\end{figure*}

The resulting sensitivities are presented in Fig.~\ref{fig:sens}.
Similarly to the case of the collection area, the presence of a cloud worsens the obtained sensitivity (in a few cases of highest energy bins the cloud-affected sensitivity is nominally reported as slightly better than in the cloudless case, however this is caused by statistical uncertainties). 
The sensitivity can be partially recovered by using either dedicated MC simulations (green curves) or image correction methods (red curves).
Notably, both approaches result in a similar sensitivity.
%In some cases (e.g. $T=0.6$, $h=9$~km a.g.l. at a few hundred GeV), the image correction method provides marginally better results than using MC simulations with a cloud. 
%This can be understood in terms of the usage 
The Hillas parametrisation of the images is the basis of the used IACT analysis. 
The regularity of the ellipse is affected by the clouds (i.e. part of the ellipse is dimmed by the light extinction), which can favour the approach based on the data correction rather than the implementation of the distortion effect in the MC simulations. 
The effect is expected to be the most pronounced at energies that have the shower maximum at similar heights as the cloud. 
We note however that the obtained improvement of the image correction method over the dedicated simulations method is relatively small, and could also be affected by the residual bias effects (the energy bias not only shifts the energies, but varying bias can also ``compress''  the statistics into some energies bins and ``dilute'' in others). 
Notably, even for the most opaque cloud simulated ($T=0.4$, $h=7$~km a.g.l.), the use of the image correction method only worsens the sensitivity by about 30\% with respect to the cloudless observations at the medium energies of a few hundred GeV. 

As expected, the usage of the additional image cleaning is degrading the performance at the lowest energies, however, the effect is smaller than for the collection area. 
This is likely related to the fact that at the lowest energies, the sensitivity is limited by the signal-to-background ratio, and the residual background is also being reduced by the additional cleaning. 
The worsening of the low-energy performance of the additional cleaning is increasing with the decreasing transmission of the cloud. 
At higher energies, using the additional cleaning improves the sensitivity and also reduces the impact of the cloud on the sensitivity. 

\section{Conclusions}
We have developed a simple geometrical model that relates the position of pixels in an IACT camera to the height responsible for most of the emission of Cherenkov photons registered by this pixel.
%Next, the geometrical model was confronted with full shower simulations (but without losing generality by applying simulations of a particular telescope) to derive small correction factors. 
The final model for the correction of cloud data was derived by applying small correction factors determined via full shower simulations.

The cloud-correction method relies on the knowledge of the vertical transmission profile of the cloud through which the Cherenkov light propagates. 
Both the current IACTs and the planned CTAO are equipped with devices capable of such measurements. 
For example, in the case of MAGIC an elastic LIDAR is used \citep{Schmuckermaier23e}, which measures the atmospheric transmission every few minutes. 
For CTAO a Raman LIDAR is planned \citep{2019arXiv190909342B}, that is sensitive to both molecular and aerosol backscatter. 
Together with FRAM (F/Photometric Robotic Atmospheric Monitor, \citealp{2021AJ....162....6E}), it will provide a full 3D determination of the aerosol extinction affecting the CTAO FoVs. 
However, the  Raman LIDAR light pulses are intense enough such that backscattered light may interfere with the CTAO telescope science observations, therefore measurements might be possible only every 20 -- 30 minutes during the repointing time of the telescopes. 
%On the other hand ceilometer \citep{2015EPJWC..8902005D} might be also used for continuous monitoring of the transmission profile, if the cloud presence is uniform and/or observations are performed at low zenith angle. 

After the correction with the above correction model, we introduced two variations in the analysis: 
%On the basis of the geometrical model, we introduced an image correction method in two incarnations: 
reusing the original image cleaning, or applying a conservative, higher image cleaning, appropriate for a particular cloud transmission. 
The proposed method was applied to full simulations of a four LST array for both variations of the image correction method. 
We have checked the effect of the cloud on the image parameters and validated if the effect can be counteracted with the proposed correction method. 
The most affected parameters: \textit{intensity}, \textit{height of the shower maximum} and \textit{length} develop a bias towards lower values induced by the cloud.
The bias is almost completely corrected with the proposed method. 
Similar to individual image parameters, the aggregated gamma/hadron separation parameter, \textit{gammaness} is also affected by the cloud presence, but can be efficiently corrected with the proposed method. 

We derived the typical performance parameters of IACTs: collection area, energy bias and resolution, angular resolution, and sensitivity.
Except for angular resolution, where the effect of a cloud is small, all performance parameters are strongly affected by more opaque clouds. 
We compared the achieved performance of the proposed image correction method with the resource-demanding approach of using dedicated MC simulations for each cloud condition. 
We found that the proposed method performs comparably to that when using the dedicated MC simulations. 
At the price of only slightly increased systematic errors, the proposed method can be applied at energies above the analysis threshold for clouds with the transmission of $\gtrsim 0.6$.
The angular and energy resolution obtained with this method is slightly worse than with dedicated simulations, but the effect of the clouds on these parameters is not dramatic. 

In the case of using an additional cleaning, the cloud correction can also be used close to the energy threshold, resulting in reduced systematic errors.  
The method is efficient in correcting most of the energy bias and provides a comparable or even slightly better sensitivity than dedicated MC simulations. 
By construction it increases the energy threshold, however, the worsening of the low-energy sensitivity is not strong as long as the cloud transmission is $\gtrsim0.6$.

The method was validated using low-zenith simulations. 
For observations at a much larger zenith angle ($\gtrsim 45^\circ$), additional factors could reduce its performance and reliability. 
In particular, the lateral distribution of the shower at higher zenith angles would result in larger spreads of the emission heights associated with the same point on the shower axis (Eq.~\ref{eq1}).
Also, for a horizontally-narrow, but vertically-broad cloud, the shower photons can skim through the edge of the cloud at high zenith not being affected by the full transmission of the cloud. 
Finally, as the proposed method relies on the measurement of the cloud transmission profile typically performed with a LIDAR, for higher zenith angle observations such devices are burdened with a lower signal-to-noise ratio.  
Nevertheless, as reported in \citet{Fruck:2022}, the elastic LIDAR used for the MAGIC telescopes was able to characterise a cloud as high as 17 km a.g.l. during observations at the zenith angle of $55^\circ$.  

We conclude that the method is valid for application in stereoscopic IACT systems, in particular in cases when slight increases of systematic uncertainties are acceptable. 
In such cases, it would allow analysis of data with severely decreased requirements of the resources (CPU time, disk memory), reducing also the analysis carbon footprint. 
A practical use case would include e.g. data taken in the presence of fast varying clouds with rather low opacity or medium/high height. 
Another recommended use case is the fast online or on-site analysis.
Such analysis cannot rely on dedicated MC simulations, but using the proposed method would allow to e.g. improve the reliability of derived fluxes for observed flares of fast transients, that need to be circulated quickly within the community. 

\section*{Acknowledgements}
This work is supported by Narodowe Centrum Nauki grant number 2019/34/E/ST9/00224.
This work was conducted in the context of the CTA LST Project. 
The authors would like to thank the internal LST and CTA Collaboration reviewers: F. Schmuckermaier, F. Di Pierro, V. de Souza and S. Einecke as well as an anonymous journal reviewer for many comments that helped to improve this manuscript. 
%% The Appendices part is started with the command \appendix;
%% appendix sections are then done as normal sections
%% \appendix

%% \section{}
%% \label{}

%% If you have bibdatabase file and want bibtex to generate the
%% bibitems, please use
%%
%%  \bibliographystyle{elsarticle-harv} 
%%  \bibliography{<your bibdatabase>}

%% else use the following coding to input the bibitems directly in the
%% TeX file.

\end{document}